\documentclass[11pt,draftclsnofoot,onecolumn,journal]{IEEEtran} 

\usepackage[english]{babel}
\usepackage[latin1]{inputenc}
\usepackage{amssymb,amstext}
\usepackage[cmex10]{amsmath}
\usepackage[ruled,vlined]{algorithm2e}
\usepackage{verbatim} 
\usepackage{float}
\usepackage[normalem]{ulem}
\usepackage{color}

\usepackage{cite}

\ifCLASSINFOpdf
   \usepackage[pdftex]{graphicx}
   \DeclareGraphicsExtensions{.pdf,.jpeg,.png}
\else
   \usepackage[dvips]{graphicx}
   \DeclareGraphicsExtensions{.eps}
\fi

\makeatletter
\def\markboth#1#2{\def\leftmark{\@IEEEcompsoconly{\sffamily}\MakeUppercase{\protect#1}}%
\def\rightmark{\@IEEEcompsoconly{\sffamily}\MakeUppercase{\protect#2}}}
\makeatother

\newtheorem{theorem}{Theorem}

\begin{document}
%
\title{\vspace{-3cm}Filter Design with Secrecy Constraints: \\
       The MIMO Gaussian Wiretap Channel}
%
%
%

\author{Hugo~Reboredo\textbf{*},~\IEEEmembership{Student Member,~IEEE,}
        Jo\~{a}o~Xavier,~\IEEEmembership{Member,~IEEE,}
        and~Miguel~R.~D.~Rodrigues,~\IEEEmembership{Member,~IEEE\vspace{-1cm}}
\thanks{\small This work was supported by Funda\c{c}\~{a}o para a Ci\^{e}ncia e Tecnologia through the research project CMU-PT/RNQ/0029/2009 and through the doctoral grant SFRH/BD/81543/2011. The work of J. Xavier is supported by Funda\c{c}\~{a}o para a Ci\^{e}ncia e a Tecnologia under Grant [PEst- OE/EEI/LA0009/2011] and Grant CMU-PT/SIA/0026/2009.\vspace{0.2cm}}
\thanks{\small H. Reboredo is with the Instituto de Telecomunica\c{c}\~{o}es and the Dept. 
de Ci\^{e}ncia de Computadores da Faculdade de Ci\^{e}ncias da Universidade do Porto, Portugal 
({\sf email: hugoreboredo@dcc.fc.up.pt}).}
\thanks{\small J. Xavier is with the Instituto de Sistemas e Rob\'otica, 
Instituto Superior T\'ecnico, ~Lisboa, ~Portugal 
~~~({\sf email: jxavier@isr.ist.utl.pt}).}
\thanks{\small M. R. D. Rodrigues is with the Department of Electronic and Electrical Engineering, University College London, United Kingdom 
~~~({\sf email: m.rodrigues@ucl.ac.uk}).}
}

\maketitle

\begin{abstract}

This paper considers the problem of filter design with secrecy constraints, where two legitimate parties (Alice and Bob) communicate in the presence of an eavesdropper (Eve), over a Gaussian multiple-input-multiple-output (MIMO) wiretap channel. This problem involves designing, subject to a power constraint, the transmit and the receive filters which minimize the mean-squared error (MSE) between the legitimate parties whilst assuring that the eavesdropper MSE remains above a certain threshold. We consider a general MIMO Gaussian wiretap scenario, where the legitimate receiver uses a linear Zero-Forcing (ZF) filter and the eavesdropper receiver uses either a ZF or an optimal linear Wiener filter. We provide a characterization of the optimal filter designs by demonstrating the convexity of the optimization problems. We also provide generalizations of the filter designs from the scenario where the channel state is known to all the parties to the scenario where there is uncertainty in the channel state. A set of numerical results illustrates the performance of the novel filter designs, including the robustness to channel modeling errors. In particular, we assess the efficacy of the designs in guaranteeing not only a certain MSE level at the eavesdropper, but also in limiting the error probability at the eavesdropper. We also assess the impact of the filter designs on the achievable secrecy rates. The penalty induced by the fact that the eavesdropper may use the optimal non-linear receive filter rather than the optimal linear one is also explored in the paper.
\end{abstract}



%
\IEEEpeerreviewmaketitle

\vspace{-0.35cm}
\section{Introduction}
%
%
%
%

\IEEEPARstart{T}{he} issues of privacy and security in wireless communication
networks have taken on an increasingly important role as these
networks continue to flourish worldwide. Traditionally, security is
viewed as an independent feature with little or no relation to the
remaining data communication tasks and, therefore, state-of-the-art
cryptographic algorithms are insensitive to the physical nature of
the wireless medium.

However, there has been more recently a renewed interest on
physical-layer security which, motivated by advances on information-theoretic
security, calls for the use of physical-layer
techniques exploiting the inherent randomness of the communications
medium to guarantee both reliable communication between two
legitimate parties as well as secure communication in the presence
of eavesdroppers.

The basis of information-theoretic security, which builds upon
Shannon's notion of perfect secrecy~\cite{Shannon1949}, was laid by
Wyner~\cite{Wyner1975} and by Csisz\'ar and
K\"orner~\cite{CsiKor1978} who proved in seminal papers that there
exist channel codes guaranteeing both robustness to transmission
errors and a certain degree of data confidentiality. In particular,
Wyner considered the wiretap channel where two legitimate users
communicate in the presence of an eavesdropper. 
Wyner characterized the rate-equivocation region of
the wiretap channel and its secrecy capacity. 
Ever since, the
computation of the secrecy capacity of a range of communications
channels has been an important research topic~\cite{LiaPooSha09}.

For example, in~\cite{LeuHell1978} the authors considered a scenario where both the main and the eavesdropper channels are additive white Gaussian noise (AWGN) channels. They showed that the secrecy capacity of such so-called Gaussian wiretap channel is equal to the difference between the main and the eavesdropper channel capacities and, therefore, confidential communications require the Gaussian main channel to have a better signal-to-noise ratio (SNR) than the Gaussian eavesdropper channel.

Motivated by the emerging wireless applications, the evaluation of the secrecy capacity of wireless fading channels with single or multiple antennas at the transmitters, receivers and/or eavesdroppers has also attracted considerable attention as well. 

Space-time signal processing techniques for secure communications over wireless links were introduced in~\cite{Her03}. The outage secrecy capacity of slow fading channels was characterized in~\cite{BarRod2006}, where it was shown that fading alone could guarantee information-theoretic security, even when the eavesdropper average SNR is higher that the legitimate receiver average SNR. In turn, the ergodic secrecy capacity of fading channels was independently characterized in~\cite{LiaPooSha08},~\cite{LiYatTrap06} and~\cite{GopLaiGam2007}. In~\cite{ParBla05} Parada and Blahut considered the secrecy capacity of several degraded fading channels. The characterization of the secrecy capacity of multiple-input-multiple-output (MIMO) channels, which represent a model for multiple-antenna channels, can be found in ~\cite{OggHas2008},~\cite{ShaLiuUlu09},~\cite{LiuBusShaPoo09} and~\cite{KhiWor10}. The computation of optimal power allocation policies and input covariances for the MIMO Gaussian wiretap channel are covered in~\cite{LiuHouShe09} and~\cite{LiPet10}, respectively.



Another key aspect in the MIMO wiretap problem is the availability of channel state information (CSI). This problem is addressed in various works under different CSI assumptions. When the CSI about the various channels is assumed to be known to all the parties, several secrecy capacity achieving schemes, based on optimal beamforming designs that leverage the general singular value decomposition (GSVD) of the main and eavesdropper channel matrices, have been proposed (e.g.~\cite{KhiWor10} and~\cite{FakSwi11}). When the CSI about the eavesdropper channel is assumed to be limited or not available, artificial noise schemes have been proposed instead~\cite{GoeNeg08},~\cite{PeiWeiWongWang2012}, where a fraction of the total power is used for reliable communication between the legitimate transmitter and the legitimate receiver and the remaining fraction of the total power is used to jam the eavesdropper. For example, the authors in~\cite{MukSwi2011} and~\cite{MukSwi09}, set up a problem whose objective is to determine the minimum transmit power necessary to guarantee a certain quality of service (QoS) between the legitimate transmitter and the legitimate receiver -- the remaining power out of the total power budget is then used to jam the eavesdropper using artificial noise type of techniques.

One key advantage of artificial noise transmission relates to the fact that the eavesdropper channel knowledge is not required. Nonetheless, the idea of transmitting artificial noise in the null space of the main channel in order to degrade the eavesdropper channel has also its limitations. On the one hand, there is an inherent trade-off between data rate and the ability to impair the eavesdropper~\cite{GoeNeg08}, so that one may not take full advantage of the spatial multiplexing ability of MIMO systems. On the other hand, if the null space of the main channel overlaps considerably with the null space of the eavesdropper channel, the artificial noise approach might lead to limited gains in security.

This paper, at the heart of the novelty of the contribution, addresses the physical-layer security problem from the estimation-theoretic
rather than the information-theoretic viewpoint. We consider the problem
of filter design with secrecy constraints in the classical MIMO wiretap
scenario consisting of two legitimate parties that communicate in
the presence of an eavesdropper, where the objective is to conceive
transmit and receive filters that, subject to a power constraint, minimize the mean-squared error
(MSE) between the legitimate parties whilst assuring that the eavesdropper MSE remains above
a certain threshold. Interestingly, this class of problems, which differs from previous approaches 
in physical-layer security in the literature (see, e.g.,~\cite{KhiWor10},~\cite{FakSwi11},~\cite{GoeNeg08},~\cite{MukSwi2011} and~\cite{MukSwi09}), represents a natural
generalization of filter design without secrecy constraints for point-to-point communications
systems (e.g.,~\cite{ScaGiaBar99},~\cite{Ges03},~\cite{PalCioLag2003},~\cite{JohUtsNos05},~\cite{PerRodVer2008},~\cite{BerPalOtt09}). 

One notable merit of this approach, in contrast to the information-theoretic work that relies on non-constructive random-coding arguments to demonstrate that there exist secrecy capacity achieving codes, is that it leads to realizable designs which can be easily implemented in practical systems. Instead, practical secrecy capacity achieving code designs are known only in some scenarios, which include: i) the main channel is noiseless and the eavesdropper channel is a binary erasure channel~\cite{SurSubBlo10},~\cite{ThaDihCal07}; ii) both channels are binary input symmetric discrete memoryless channels (DMC) and the eavesdropper channel is degraded with respect to the main channel -- where polar codes are used~\cite{MahVar11},~\cite{KoyGam12}; and iii) the eavesdropper is constrained combinatorially~\cite{OzaWyn84}.

Nonetheless, it is relevant to pause to reflect on the operational relevance of this new metric, in view of the fact that it is the norm, in the information-theoretic security literature, to use equivocation rather than MSE to measure security. In fact, the use of the MSE in \emph{lieu} of equivocation does not guarantee perfect information-theoretic security in the sense of~\cite{Shannon1949},~\cite{Wyner1975} and ~\cite{CsiKor1978}. We view the design of the filters based on the MSE criteria as a means to provide additional confusion in a communications system. 

The rationale of the new design approach is then based on the fact that some applications require a MSE below a certain level to function properly, so that this approach would impair further the performance of the eavesdropper by imposing a threshold on its MSE level. Note also that the bit error rate (BER), which is a very important figure of merit in a communications system, is typically monotonically increasing with the MSE, so that a threshold on the MSE may also translate into a threshold in the BER.


One particular scenario that suits this design approach relates to wireless broadcasting where a service provider provides different services, e.g. different video streams, to different users/subscribers (see Figure~\ref{fig:broadcast}). Here, the service provider (the legitimate transmitter) needs to guarantee that a user that has subscribed to the service (the legitimate receiver) has access to a high quality version of the video stream whereas a user that has not subscribed to the service (the so-called eavesdropper) has only access to a very poor quality version of the video stream. The use of a distortion metric, such as the MSE or the BER, instead of equivocation, is then entirely appropriate for this class of applications, offering an alternative to the cryptographic methods used by Content Access (CA) systems~\cite{MacQui95},~\cite{Noo03},~\cite{GalTom05}.

It turns out thus that the filter design with secrecy constraints problem is to be understood broadly as a filter design problem with distortion constraints. However, in order to connect this work with the large body of work of physical- and information-theoretic security whose overarching aim is to impair the eavesdropper, we -- in a somewhat abusive use of language -- use the notion secrecy rather than distortion.

This paper is structured as follows: 
Section \ref{problem} defines
the problem. Section \ref{ZF_filters} considers the design of the transmit filter
when ZF filters are used at both the legitimate and the eavesdropper receivers. 
In turn, Section \ref{wiezf_filters} considers the design of the transmit filter when the eavesdropper 
uses an optimal linear filter while the legitimate receiver is restricted to the use of a ZF receive filter. Section \ref{fading} provides some generalizations of the problem of filter design with secrecy constraints, from the scenario where the state of the channels is known exactly to all the parties (i.e., the legitimate transmitter, the legitimate receiver and the eavesdropper) to the scenario where there is uncertainty in the channel state.
Section \ref{results} shows various numerical results to illustrate the
impact of the filter designs on both the reliability and security criteria, evaluating, not only the MSE, but also the bit error rate and the achievable secrecy rates yielded by the designs. The main contributions of the manuscript are
summarized in Section \ref{conclusions}.

\subsection{Notation}
\label{notation}

We use the following notation: boldface upper-case letters denote matrices or column vectors (${\bf X}$) and italics denote scalars ($x$); the context defines whether the quantities are deterministic or random. The notation ${\bf M} \succ {\bf 0}$ is used to denote a positive definite matrix and ${\bf M} \succeq {\bf 0}$ denotes a positive semidefinite matrix. The symbol ${\bf I}$ represents the identity matrix. The operators
$\|\cdot\|^2$, ${\sf tr}\left\{\cdot\right\}$ and $\nabla$ represent the $l_2$-norm, the trace operator and the gradient operator, respectively. The operators $\left(\cdot\right)^{\dagger}$ and $\left(\cdot\right)^{+}$ denote the Hermitian transpose operator and the Pseudo-Inverse operator, respectively. The operator $\mathcal{E}\left(\cdot\right)$ represents the expectation. $\mathcal{CN}\left(\mu,{\bf \Sigma}\right)$ denotes a circularly symmetric complex Gaussian random vector with mean $\mu$ and covariance ${\bf \Sigma}$.

\section{Problem Statement}
\label{problem}

\begin{figure}[!htbp]
    \begin{center}
        \mbox{\includegraphics[width=4.5in]{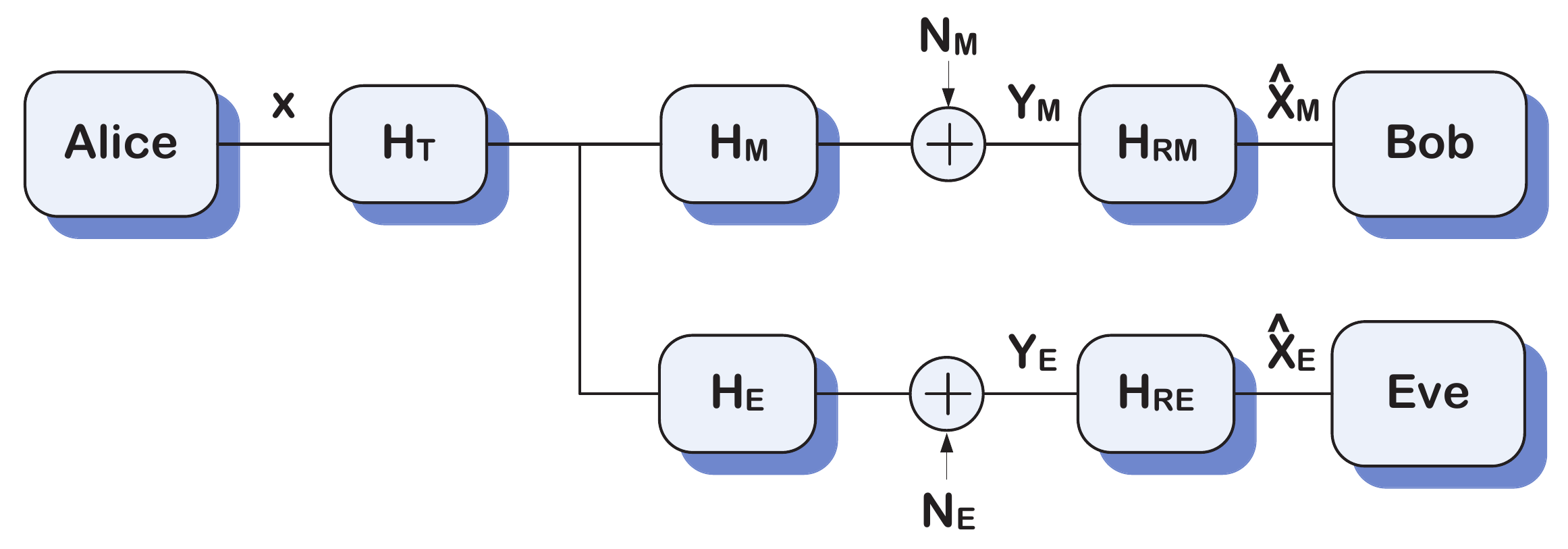}}
        \vspace{-0.5cm}
    \caption{MIMO Gaussian wiretap channel model.}
    \label{fig:model}
    \end{center}
    \label{model}
\end{figure}

We consider a communications scenario where a legitimate user, say
Alice, communicates with another legitimate user, say Bob, in the
presence of an eavesdropper, Eve (see Figure \ref{fig:model}).

Bob and Eve observe the output of the MIMO channels given, respectively, by:

\begin{equation}
\label{bob_channel}
{\bf Y}_M = {\bf H}_M {\bf H}_T {\bf X} + {\bf N}_M
\end{equation}
\begin{equation}
\label{eve_channel}
{\bf Y}_E = {\bf H}_E {\bf H}_T {\bf X} + {\bf N}_E
\end{equation}
where ${\bf Y}_M \in \mathbb{C}^{n_M}$ and ${\bf Y}_E \in \mathbb{C}^{n_E}$ are the vectors of receive
symbols, ${\bf X} \in \mathbb{C}^{m}$ is the vector of independent,
zero-mean and unit-variance transmit symbols, and ${\bf N}_M \in \mathbb{C}^{n_M}$ and ${\bf N}_E \in \mathbb{C}^{n_E}$ are 
circularly symmetric complex Gaussian random vector with zero mean and
identity covariance matrix\footnote{\scriptsize The models in \eqref{bob_channel} and in \eqref{eve_channel} follow from the more general models ${\bf \tilde{Y}}_M = {\bf \tilde{H}}_M {\bf H}_T {\bf X} + {\bf \tilde{N}}_M$ and ${\bf \tilde{Y}}_E = {\bf \tilde{H}}_E {\bf H}_T {\bf X} + {\bf \tilde{N}}_E$, respectively, where ${\bf \tilde{N}}_M$ and ${\bf \tilde{N}}_E$ are circularly symmetric complex Gaussian random vectors with mean $\mathcal{E}\left( {\bf \tilde{N}}_M\right) = 0$ and $\mathcal{E}\left( {\bf \tilde{N}}_E\right) = 0$, and covariance matrices $\mathcal{E}\left({\bf \tilde{N}}_M {\bf \tilde{N}}_M^{\dag}\right) = {\bf \Sigma}_{N_M}$ and $\mathcal{E}\left({\bf \tilde{N}}_E {\bf \tilde{N}}_E^{\dag}\right) = {\bf \Sigma}_{N_E}$, respectively, by using pre-whitening filters i.e., ${\bf Y}_M = {\bf \Sigma}_{N_M}^{-1/2}{\bf \tilde{Y}}_M$ $= {\bf \Sigma}_{N_M}^{-1/2}{\bf \tilde{H}}_M {\bf H}_T {\bf X} + {\bf \Sigma}_{N_M}^{-1/2}{\bf \tilde{N}}_M$ $= {\bf H}_M {\bf H}_T {\bf X} + {\bf N}_M$ and ${\bf Y}_E = {\bf H}_E {\bf H}_T {\bf X} + {\bf N}_E$. These transformations are information lossless~\cite{Pal03}.}.
The $n_M \times m$ matrix ${\bf H}_M$ and the $n_E \times m$ matrix ${\bf
H}_E$ contain the deterministic gains from each main and eavesdropper channel input
to each main and eavesdropper channel output, respectively. The $m \times m$ matrix ${\bf H}_T$
represents Alice's transmit filter.

\begin{figure}[!htbp]
    \begin{center}
        \mbox{\includegraphics[width=5.70in]{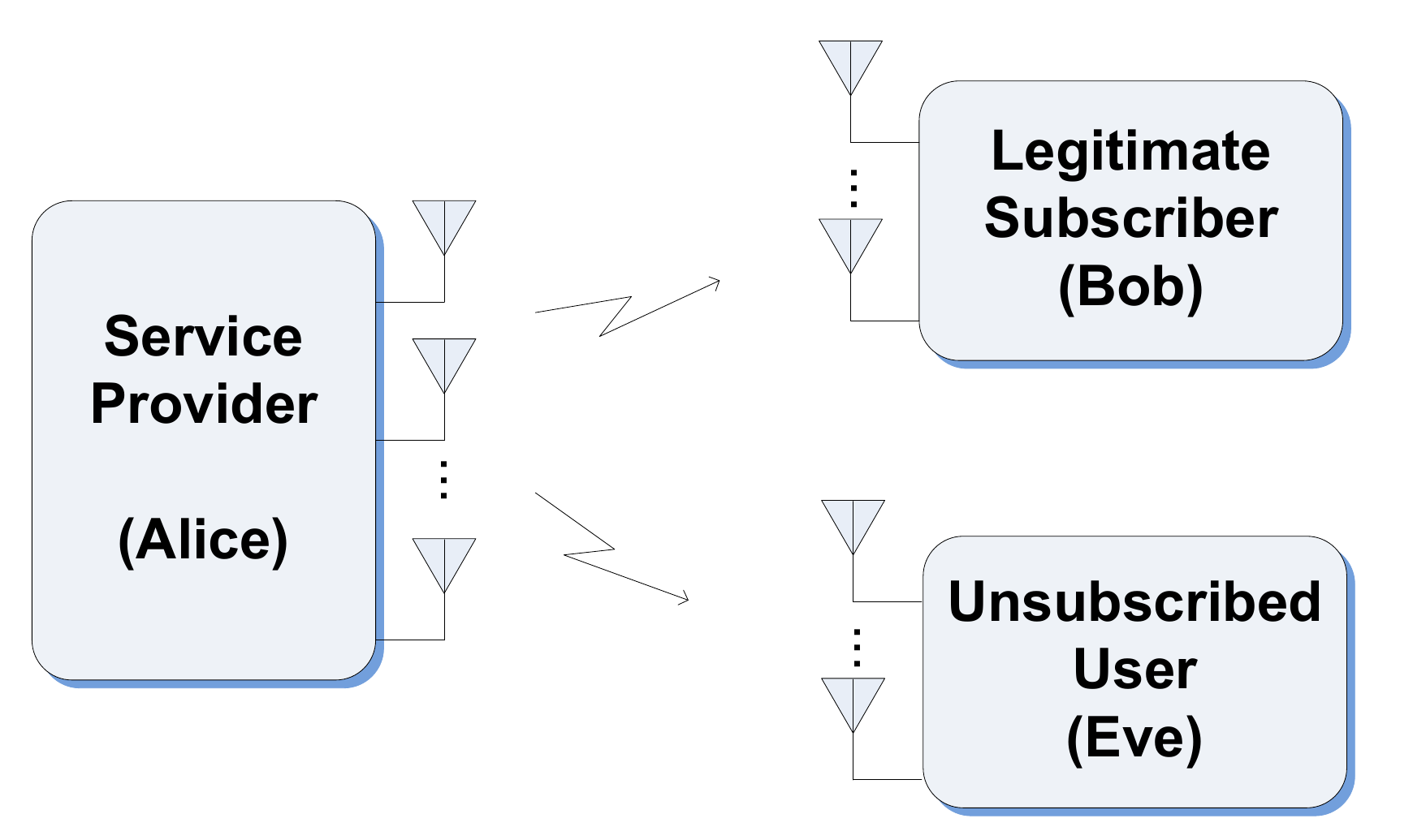}}
        \vspace{-0.5cm}
   \caption{A possible application scenario of the problem of filter design with secrecy constraints: "Secure" video broadcasting.}
    \label{fig:broadcast}
    \end{center}
    \label{broadcast}
\end{figure}

We assume that ${\bf H}_M{\bf H}_T$ and ${\bf H}_E{\bf H}_T$ are full column rank, which implies that $n_M \geq m$ and $n_E \geq m$. This is necessary to guarantee the existence of some solutions. We further assume that, in a realistic scenario, the channel matrices ${\bf H}_M$ and ${\bf H}_E$ are not a multiple of each other. We also assume that the channel state is known by all the parties, i.e. Alice, Bob and Eve have perfect knowledge about the channel matrices ${\bf H}_M$ and ${\bf H}_E$. This is often a common assumption in the physical layer security literature (see e.g. \cite{BarRod2006} and \cite{DonHaPetPoor10}). The assumption that the legitimate receiver knows the state of the main channel and the eavesdropper receiver knows the state of the wiretap channel is realistic, because the receivers can always estimate the channels in slow fading conditions. The assumption that the transmitter knows the state of the main channel and, more importantly, the wiretap channel or that the legitimate receiver knows the state of the wiretap channel and the eavesdropper knows the state of the main channel can be justified in wireless networks where the eavesdropper is another network active user (e.g. in the scenario of Figure~\ref{fig:broadcast}). In particular, in time division duplex (TDD) environments Alice can estimate the state of Bob's and Eve's channels and inform the receivers accordingly. However, we will also generalize the framework to incorporate possible channel uncertainties in the sequel.

Bob's and Eve's estimate of the vector of input symbols are, respectively, given
by:
\begin{equation}
{ {\bf \hat{X}}_M = {{\bf H}_R}_M {\bf Y}_M}
\end{equation}
\begin{equation}
{ {\bf \hat{X}}_E = {{\bf H}_R}_E {\bf Y}_E}
\end{equation}
where the $m \times n_M$ matrix ${{\bf H}_R}_M$ and the $m \times
n_E$ matrix ${{\bf H}_R}_E$ represent Bob's and Eve's receive
filters, respectively.

In this setting, we take, as a performance metric, the MSE between the estimate of the input vector ${\bf \hat{X}}$ and the true
input vector ${\bf X}$ given by:
\begin{equation}
{ {\sf MSE} = \mathcal{E} \left[\|{\bf X} - {\bf \hat{X}}\|^2\right]}
\end{equation}

The objective is to design, for specific receive filter choices, the transmit filter that solves the optimization problem:
\begin{equation}\label{opt1a}
{ \min {\sf MSE_{M}} = \mathcal{E} \left[\|{\bf X} - {\bf
\hat{X}}_M\|^2\right]}
\end{equation}
subject to the security constraint: 
\begin{equation}\label{opt1b}
{ {\sf MSE_{E}} = \mathcal{E} \left[\|{\bf X} - {\bf \hat{X}}_E\|^2\right] \geq \gamma}
\end{equation}
where $\gamma$ represents an MSE threshold, and to the total power constraint:
\begin{equation}\label{opt1c}
{ {\sf tr} \left\{{\bf H}_T {\bf H}_T^\dag\right\} \leq P_{avg}}
\end{equation}
where $P_{avg}$ represents the available power.

We restrict our attention to two specific design scenarios: i) the situation where both the legitimate  receiver and the eavesdropper receiver are constrained to obey ZF constraints; and ii) the situation where the legitimate receiver uses a ZF filter whereas the eavesdropper receiver uses the optimal linear Wiener filter. For these receiver filter choices, the optimization problem in \eqref{opt1a} -- \eqref{opt1c} is convex thus enabling the characterization of optimal designs; for other receiver filter choices, and to the best of our knowledge, the optimization problem in \eqref{opt1a} -- \eqref{opt1c} is only convex for special scenarios, e.g. the degraded parallel Gaussian wiretap channel, or the degraded MIMO wiretap channel (see \cite{RodAlm2008} and \cite{RebRod2010vtc})\footnote{\scriptsize We prove the convexity of the filter design with secrecy constraints optimization problem by using the change of variables ${\bf Z} = \left( {\bf H}_T {\bf H}_T^\dag \right)^{-1}$. This change of variables leads to convex objective functions as well as convex feasible regions when both the legitimate receiver and the eavesdropper receiver use ZF filters  (see \eqref{opt2a}, \eqref{opt2b} and \eqref{opt2c}) and when the legitimate receiver uses a ZF filter but the eavesdropper receiver uses a Wiener filter (see \eqref{opt_1}, \eqref{opt_2} and \eqref{opt_3}). However, such a change of variables does not lead immediately to a convex optimization problem when both the legitimate receiver and the eavesdropper receiver adopt the Wiener filter (the feasible region is still convex but the objective function is concave rather than convex). Thus -- with the exception of~\cite{RodAlm2008} and \cite{RebRod2010vtc} -- it is not entirely clear whether other change of variables lead to a convex optimization problem in such a case.}. 

We recognize that our formulation assumes the so-called eavesdropper to perform a certain linear action whereas the traditional information-theoretic formulation -- in view of the fact that it is based on the equivocation metric -- does not assume the eavesdropper to perform any specific operation. However, 
in the scenario where the eavesdropper is another user of the network as in Figure~\ref{fig:broadcast},
it seems appropriate to assume a certain action by this user. We also recognize the fact that a more sophisticated eavesdropper would possibly leverage nonlinear techniques to estimate the information. This issue is also discussed in the sequel. 

It is also important to note that, and in contrast to the artificial noise approach in~\cite{GoeNeg08},~\cite{PeiWeiWongWang2012}, ~\cite{MukSwi2011},~\cite{MukSwi09} and~\cite{Swi09}, our filter design approach does not impose a limitation on the ability of transmitting information along all the dimensions that the MIMO channel has to offer and, therefore, we can expect to achieve higher data rates. However, by imposing a threshold on the eavesdropper MSE we may also naturally constraint the performance of the main channel.

\section{Zero Forcing Filters at the Receivers}
\label{ZF_filters}

We now  consider the scenario where both the legitimate receiver and the eavesdropper receiver use ZF filters, thus obeying the ZF constraints given by:
\begin{align}
&{  { \bf H}_{RM} { \bf H}_M { \bf H}_T = { \bf I} \label{ZF_constraint1}} \\
&{  { \bf H}_{RE} { \bf H}_E { \bf H}_T = { \bf I} \label{ZF_constraint2}}
\end{align}
The rationale for including the ZF constraints in \eqref{ZF_constraint1} and \eqref{ZF_constraint2} is to eliminate crosstalk between the various streams (e.g.~\cite{HoCreSte92}). Note also that the performance of ZF linear receivers is equivalent to that of optimal Wiener linear receivers in the regime of high SNR. Yet, one may still argue that a eavesdropper will always adopt the optimal linear receive filter (or the optimal non-linear receive filter), rather than the sub-optimal ZF receive filter. These particular cases will be addressed in Sections \ref{wiezf_filters} and \ref{conclusions}.

\subsection{Optimal Receive Filters}
\label{rx_filters_zf}

Let us consider the design of the receive filters. Bob uses the receive filter that, for any fixed transmit filter ${ \bf H}_T$, minimizes:
\begin{equation}\label{m1}
{ {\sf MSE_{M}} = \mathcal{E} \left[\|{\bf X} - {\bf
\hat{X}}_M\|^2\right] = \mathcal{E} \left[\|{\bf X} - {{\bf H}_R}_M
{\bf Y}_M\|^2\right]}
\end{equation}
subject to the ZF constraint in \eqref{ZF_constraint1} and Eve uses the receive filter that, for any fixed transmit filter ${ \bf H}_T$, minimizes:
\begin{equation}\label{m2}
{ {\sf MSE_{E}} = \mathcal{E} \left[\|{\bf X} - {\bf
\hat{X}}_E\|^2\right] = \mathcal{E} \left[\|{\bf X} - {{\bf H}_R}_E
{\bf Y}_E\|^2\right]}
\end{equation}
subject to the ZF constraint in \eqref{ZF_constraint2}.

In particular, the receive filters, which follow immediately from \eqref{ZF_constraint1} and \eqref{ZF_constraint2},  are given by~\cite{Pal03}:
\begin{align}
& { {{ \bf H}_{R_M}^{*}} =  \left({ \bf H}_M{ \bf H}_T\right)^+ = \left({ \bf H}_T^\dag { \bf H}_M^\dag { \bf H}_M{ \bf H}_T\right)^{-1}{ \bf H}_T^\dag { \bf H}_M^\dag \label{Rx_m}}\\
& { {{ \bf H}_{R_E}^{*}} =  \left({ \bf H}_E{ \bf H}_T\right)^+ = \left({ \bf H}_T^\dag { \bf H}_E^\dag { \bf H}_E{ \bf H}_T\right)^{-1}{ \bf H}_T^\dag { \bf H}_E^\dag \label{Rx_e}}
\end{align}

The MSEs in the main and eavesdropper channels, upon substituting \eqref{Rx_m} and \eqref{Rx_e} in \eqref{m1} and \eqref{m2}, respectively, are then given by:
\begin{align}
& {   {{\sf MSE_M}} = \mathcal{E} \left[\|{\bf X} - {{ \bf H}_{R_M}^{*}} {\bf Y}_M\|^2\right] 
= {\sf tr}\left\{ \left({ \bf H}_T^\dag { \bf H}_M^\dag { \bf H}_M{ \bf H}_T\right)^{-1}\right\}} \label{MSE_M}\\
& {   {{\sf MSE_E}} = \mathcal{E} \left[\|{\bf X} - {{ \bf H}_{R_E}^{*}} {\bf Y}_E\|^2\right] 
= {\sf tr}\left\{ \left({ \bf H}_T^\dag { \bf H}_E^\dag { \bf H}_E{ \bf H}_T\right)^{-1}\right\}} \label{MSE_E}
\end{align}

\subsection{Optimal Transmit Filter}
\label{tx_filter_zf}

In view of \eqref{MSE_M} and \eqref{MSE_E}, the form of the optimal transmit filter corresponds to the solution of the optimization problem:  
\begin{equation}\label{opt2a}
{ \min_{{\bf H}_T}~~{\sf tr}\left\{ \left({ \bf H}_T^\dag { \bf H}_M^\dag { \bf H}_M{ \bf H}_T\right)^{-1}\right\} }
\end{equation}
subject to the constraints:
\begin{equation}\label{opt2b}
{ {\sf tr}\left\{ \left({ \bf H}_T^\dag { \bf H}_E^\dag { \bf H}_E{ \bf H}_T\right)^{-1}\right\} \geq \gamma }
\end{equation}
\begin{equation}\label{opt2c}
{ {\sf tr}\left\{ { \bf H}_T { \bf H}_T^\dag \right\} \leq P_{avg} }
\end{equation}
and ${ \bf H}_T { \bf H}_T^\dag \succ {\bf 0}$ (Note that ${ \bf H}_T { \bf H}_T^\dag \succ {\bf 0}$, because ${\bf H}_M{\bf H}_T$ and ${\bf H}_E{\bf H}_T$  are full column rank by assumption). Note that -- due to the channel knowledge assumptions -- the legitimate transmitter, the legitimate receiver and the eavesdropper can all set up this optimization problem in order to determine the transmit filter and hence the receive filters via \eqref{Rx_m} and \eqref{Rx_e}.

It is now possible to reduce this optimization problem to a standard convex optimization problem by adopting the change of variables ${ \bf Z} = \left({ \bf H}_T { \bf H}_T^\dag \right)^{-1} $, thereby paving the way to the characterization of the optimal transmit filter.

The following Theorem, which stems directly from the
Karush-Kuhn-Tucker optimality conditions~\cite{Boyd04}, defines the form of the optimal
transmit filter.

\vspace{0.20cm}
\begin{theorem}
\label{optimaltx_zf}
Assume that the legitimate transmitter, the legitimate receiver and the eavesdropper know the exact channel matrices ${\bf H}_M$ and ${\bf H}_E$. Assume also that the legitimate receiver and the eavesdropper receiver use ZF filters. Then, an optimal transmit filter that solves the optimization problem in \eqref{opt2a} -- \eqref{opt2c} is, without loss of generality, given by:
\[ {  {{\bf H}_T^{\ast}}=} \left\{
\begin{array}{l l}
{  \sqrt{\frac{P_{avg}}{{\sf tr} \left\{\left({ \bf H}_M^\dag { \bf H}_M \right)^{-\frac{1}{2}}\right\}}} \left({ \bf H}_M^\dag { \bf H}_M \right)^{-\frac{1}{4}},}\\ 
{  \qquad \frac{{\sf tr} \left\{\left({ \bf H}_M^\dag { \bf H}_M \right)^{-\frac{1}{2}}\right\}}{P_{avg}} {\sf tr} \left\{\left({ \bf H}_E^\dag { \bf H}_E \right)^{-1}\left({ \bf H}_M^\dag { \bf H}_M \right)^{\frac{1}{2}} \right\} > \gamma} \\ 
\\
{  \sqrt{\frac{P_{avg}}{{\sf tr} \left\{\left[\left[{ \bf H}_M^\dag { \bf H}_M \right]^{-1} - \nu \left[{ \bf H}_E^\dag { \bf H}_E \right]^{-1}\right]^{\frac{1}{2}}\right\}}} \left[\left[{ \bf H}_M^\dag { \bf H}_M \right]^{-1} - \nu \left[{ \bf H}_E^\dag { \bf H}_E \right]^{-1}\right]^{\frac{1}{4}},}\\
{  \qquad \frac{{\sf tr} \left\{\left({ \bf H}_M^\dag { \bf H}_M \right)^{-\frac{1}{2}}\right\}}{P_{avg}} {\sf tr} \left\{\left({ \bf H}_E^\dag { \bf H}_E \right)^{-1}\left({ \bf H}_M^\dag { \bf H}_M \right)^{\frac{1}{2}} \right\} \leq \gamma} 
\end{array}
\right. \]
where the value of the Lagrange multiplier $\nu$ is such that:
\begin{multline}
\label{nu}
{  {\sf tr} \left\{ \left({ \bf H}_E^\dag { \bf H}_E \right)^{-1}\left(\left({ \bf H}_M^\dag { \bf H}_M \right)^{-1} - \nu \left({ \bf H}_E^\dag { \bf H}_E \right)^{-1}\right)^{-1/2}\right\}~\times} \\
{  \times~{\sf tr} \left\{\left( \left({ \bf H}_M^\dag { \bf H}_M \right)^{-1} - \nu \left({ \bf H}_E^\dag { \bf H}_E \right)^{-1}\right)^{1/2}\right\}  = \gamma \cdot P_{avg}}
\end{multline}

\end{theorem}
Note that the right multiplication of the transmit filter in Theorem~\ref{optimaltx_zf} by any unitary matrix produces another optimal filter.
\vspace{0.50cm}
\begin{IEEEproof}
By considering the change of variables ${ \bf Z} = \left({ \bf H}_T { \bf H}_T^\dag \right)^{-1} $ 
it is possible to rewrite the optimization problem in \eqref{opt2a} -- \eqref{opt2c} as follows:
\begin{equation}
\label{zf_prob_1a}
{ \min_{{\bf Z}}~~{\sf tr}\left\{ \left({ \bf H}_M^\dag { \bf H}_M \right)^{-1} { \bf Z} \right\}} 
\end{equation}
subject to the constraints ${\sf tr}\left\{ \left({ \bf H}_E^\dag { \bf H}_E \right)^{-1} { \bf Z} \right\} \geq \gamma$, ${\sf tr}\left\{ { \bf Z}^{-1} \right\} \leq P_{avg}$, 
and ${\bf Z} \succ {\bf 0}$. Note that this represents a standard convex optimization problem, so that the solution follows directly from the Karush-Kuhn-Tucker optimality conditions~\cite{Boyd04}.

The Lagrangian of the optimization problem is given by:
\begin{align}
\label{lagr1}
{  \mathfrak{L} \left( {\bf Z},\nu,\mu \right)}
& {  = {\sf tr}\left\{\left({ \bf H}_M^\dag { \bf H}_M \right)^{-1}{\bf Z}\right\} + \nu \left(\gamma - {\sf tr}\left\{\left({ \bf H}_E^\dag { \bf H}_E \right)^{-1}{\bf Z}\right\} \right) + \mu \left({\sf tr}\left\{{\bf Z}^{-1}\right\} - P_{avg} \right)}
\end{align}
where $\nu$ and $\mu$ are the Lagrange multipliers associated with the problem constraints. The Karush-Kuhn-Tucker optimality conditions are given by:
\begin{align}\label{kkt1a}
{  \nabla_{\bf Z} \mathfrak{L} \left( {\bf Z},\nu,\mu \right) ~=~ \left({ \bf H}_M^\dag { \bf H}_M \right)^{-1} - \nu\left({ \bf H}_E^\dag { \bf H}_E \right)^{-1} - \mu {\bf Z}^{ -2} = 0}
\end{align}
\begin{align}
{ \nu \left[ {\sf tr}\left\{ \left({ \bf H}_E^\dag { \bf H}_E \right)^{-1}{\bf Z}  \right\} -\gamma \right] = 0,~~~ \nu \geq 0} \\
{ \mu \left[ P_{avg} - {\sf tr}\left\{ {\bf Z}^{-1}\right\}\right] = 0,~~~ \mu \geq 0}
\end{align}
and ${\bf Z} \succ {\bf 0}$, ${\sf tr}\left\{ \left({ \bf H}_E^\dag { \bf H}_E \right)^{-1} { \bf Z} \right\} \geq \gamma$, ${\sf tr}\left\{ { \bf Z}^{-1} \right\} \leq P_{avg}$.

The Karush-Kuhn-Tucker optimality conditions reveal that the solution of this problem exhibits two distinct regimes only: i) the regime where the secrecy constraint is not active ($\nu = 0$); and ii) the regime where the secrecy constraint is met with equality ($\nu > 0$)\footnote{\scriptsize In each case the power constraint is met with equality i.e., $\mu > 0$. Note that a scenario where the $\mu = 0$ would require either the channel matrices to be a multiple of each other ($\nu > 0$ and $\mu = 0$), or ${ \bf H}_M^\dag { \bf H}_M = { \bf 0}$ ($\nu = 0$ and $\mu = 0$).}.

When $\nu = 0$, then \eqref{kkt1a} reduces to:
\begin{align}
&{ \left({ \bf H}_M^\dag { \bf H}_M \right)^{-1} - \mu{\bf Z}^{-2} = 0}
\end{align}
and the optimal solution is given by:
\begin{equation}
{ 	{\bf Z}^{\ast}=\frac{ {\sf tr} \left\{\left({ \bf H}_M^\dag { \bf H}_M \right)^{-1/2}\right\}}{P_{avg}}\left({ \bf H}_M^\dag { \bf H}_M \right)^{1/2} }
\end{equation}
 
This solution is valid if and only if:
\begin{align}
{  \frac{{\sf tr} \left\{\left({ \bf H}_M^\dag { \bf H}_M \right)^{-1/2}\right\}}{P_{avg}} {\sf tr} \left\{\left({ \bf H}_E^\dag { \bf H}_E \right)^{-1}\left({ \bf H}_M^\dag { \bf H}_M \right)^{1/2} \right\} > \gamma}
\end{align}

On the other hand, when $\nu >0$, then \eqref{kkt1a} reduces to:
\begin{align}
&{ \left({ \bf H}_M^\dag { \bf H}_M \right)^{-1} - \nu\left({ \bf H}_E^\dag { \bf H}_E \right)^{-1} - \mu {\bf Z}^{ -2} = 0}
\end{align}
and the optimal solution is given by:
\begin{align}
& { {\bf Z}^{\ast}= \frac{ {\sf tr} \left\{\left[\left({ \bf H}_M^\dag { \bf H}_M \right)^{-1} - \nu\left({ \bf H}_E^\dag { \bf H}_E \right)^{-1}\right]^{\frac{1}{2}}\right\}}{P_{avg}} \left[\left({ \bf H}_M^\dag { \bf H}_M \right)^{-1} - \nu\left({ \bf H}_E^\dag { \bf H}_E \right)^{-1}\right]^{-\frac{1}{2}}}
\end{align}
This solution is valid if and only if:
\begin{align}
{  \frac{{\sf tr} \left\{\left({ \bf H}_M^\dag { \bf H}_M \right)^{-1/2}\right\}}{P_{avg}} {\sf tr} \left\{\left({ \bf H}_E^\dag { \bf H}_E \right)^{-1}\left({ \bf H}_M^\dag { \bf H}_M \right)^{1/2} \right\} \leq \gamma}
\end{align}

\end{IEEEproof}

Note that the optimal transmit filter obeys a simple operational interpretation. In the regime where the secrecy constraint is inactive, i.e.:
\begin{align}
{  \frac{{\sf tr} \left\{\left({ \bf H}_M^\dag { \bf H}_M \right)^{-1/2}\right\}}{P_{avg}} {\sf tr} \left\{\left({ \bf H}_E^\dag { \bf H}_E \right)^{-1}\left({ \bf H}_M^\dag { \bf H}_M \right)^{1/2} \right\} > \gamma}
\end{align}
which typically occurs for low available powers, the filter performs two simple operations: i) conversion of the main channel (i.e. ${ \bf H}_M^\dag { \bf H}_M$) into a set of parallel independent channels whose power gains correspond to the eigenvalues of the matrix ${ \bf H}_M^\dag { \bf H}_M$; and ii) power allocation, by dividing the total power inversely proportionally to the power gains of the set of parallel channels. This solution corresponds to the solution in~\cite{Pal03}. 

In contrast, in the regime where the secrecy constraint is active, i.e.:
\begin{align}
{  \frac{{\sf tr} \left\{\left({ \bf H}_M^\dag { \bf H}_M \right)^{-1/2}\right\}}{P_{avg}} {\sf tr} \left\{\left({ \bf H}_E^\dag { \bf H}_E \right)^{-1}\left({ \bf H}_M^\dag { \bf H}_M \right)^{1/2} \right\} \leq \gamma}
\end{align}
which typically occurs for high available powers, the filter can be seen to perform the operations: i) conversion of an equivalent channel (i.e. $\left({ \bf H}_M^\dag { \bf H}_M\right)^{-1}  - \nu \left({ \bf H}_E^\dag { \bf H}_E \right)^{-1}$) into a set of parallel independent channels whose power gains correspond to the eigenvalues of the matrix $\left({ \bf H}_M^\dag { \bf H}_M\right)^{-1}  - \nu \left({ \bf H}_E^\dag { \bf H}_E \right)^{-1}$ and; ii) power allocation, by dividing the total power inversely proportionally to the power gains of the set of parallel channels. This result, which is based on the equivalent channels (rather than on the main channel), immediately generalizes the result in~\cite{Pal03}.

Note also that, in the scenario where both receivers use ZF filters the power constraint is always active, i.e. the transmitter uses all the available power. We will observe in the sequel that this is not the case in other scenarios.

\subsection{Computational Procedure}
\label{algo_zf}

The computation of the optimal transmit filter embodied in Theorem \ref{optimaltx_zf}
requires finding the solution of the non-linear equation in \eqref{nu}, in order
to determine the value of the Lagrange multiplier $\nu$. We shall now put forth
a simpler procedure to design the optimal transmit filter and hence the receive filters via \eqref{Rx_m} and \eqref{Rx_e}, based on the dual
of the optimization problem.

Consider again the Lagrangian of the optimization problem in \eqref{lagr1}. 
Consider also the dual function of the optimization problem in \eqref{zf_prob_1a}:
\begin{align}
{ \mathfrak{L} \left(\nu,\mu \right)}
& { = \inf_{{\bf Z} \succeq {\bf 0}} \mathfrak{L} \left( {\bf Z},\nu,\mu \right)}
\end{align}
where $\nu \geq 0$ and $\mu \geq 0$. It is straightforward to show that the dual
function reduces to:
\[{  \mathfrak{L} \left(\nu,\mu \right) = }\left\{
\begin{array}{l l}
{  2\sqrt{\mu}~~{\sf tr} \left\{\left(\left({ \bf H}_M^\dag { \bf H}_M \right)^{-1} - \nu\left({ \bf H}_E^\dag { \bf H}_E \right)^{-1}\right)^{\frac{1}{2}}\right\} - \mu P_{avg} + \nu \gamma,}\\ 
{  \qquad \qquad ~~~~~\left(\left({ \bf H}_M^\dag { \bf H}_M \right)^{-1} - \nu\left({ \bf H}_E^\dag { \bf H}_E \right)^{-1}\right) \geq 0} \\
{  -\infty, \qquad \qquad \qquad \qquad \qquad \qquad \qquad \text{otherwise}}
\end{array}
\right. \]

The dual problem of the optimization problem in \eqref{zf_prob_1a} is now given by:
\begin{align}
\label{dual}
& {  \max_{\mu, \nu} ~~  2\sqrt{\mu}~{\sf tr} \left\{\left(\left({ \bf H}_M^\dag { \bf H}_M \right)^{-1} - \nu\left({ \bf H}_E^\dag { \bf H}_E \right)^{-1}\right)^{\frac{1}{2}}\right\} -~\mu P_{avg} + \nu \gamma}
\end {align}
subject to $\nu \geq 0$, $\mu \geq 0$ and $\left(\left({ \bf H}_M^\dag { \bf H}_M \right)^{-1} - \nu\left({ \bf H}_E^\dag { \bf H}_E \right)^{-1}\right) \succeq {\bf 0}$.
We can now employ a two step procedure to express the solution of this optimization problem:
i) optimization over $\mu$ for a fixed $\nu$; 
~~ii) optimization over $\nu$ for the optimal $\mu$. It is straightforward to show that the optimal value of $\mu$, for a fixed $\nu$, is given by:
\begin{equation}
\label{mu}
{  \mu = \frac{1}{{P_{avg}}^2} \left({\sf tr} \left\{\left(\left({ \bf H}_M^\dag { \bf H}_M \right)^{-1} - \nu\left({ \bf H}_E^\dag { \bf H}_E \right)^{-1}\right)^{\frac{1}{2}}\right\}\right)^2}
\end{equation}
Consequently, the dual optimization problem reduces to:
\begin{equation}
{  \max_{\nu} \frac{1}{P_{avg}} \left({\sf tr} \left\{\left(\left({ \bf H}_M^\dag { \bf H}_M \right)^{-1} - \nu\left({ \bf H}_E^\dag { \bf H}_E \right)^{-1}\right)^{\frac{1}{2}}\right\}\right)^2 + \nu \gamma}
\end {equation}
subject to $\nu \geq 0$ and $\left(\left({ \bf H}_M^\dag { \bf H}_M \right)^{-1} - \nu\left({ \bf H}_E^\dag { \bf H}_E \right)^{-1}\right) \succeq {\bf 0}$ or, equivalently:
\begin{equation}
\label{dual_final1}
{  \max_{\nu} \frac{1}{P_{avg}} \left({\sf tr} \left\{\left(\left({ \bf H}_M^\dag { \bf H}_M \right)^{-1} - \nu\left({ \bf H}_E^\dag { \bf H}_E \right)^{-1}\right)^{\frac{1}{2}}\right\}\right)^2 + \nu \gamma}
\end{equation}
subject to:
\begin{equation}
\label{dual_final2}
{  0 \leq \nu \leq \lambda_{min}\left(\left({ \bf H}_E^\dag { \bf H}_E \right)^{\frac{1}{2}} \left({ \bf H}_M^\dag { \bf H}_M \right)^{-1} \left({ \bf H}_E^\dag { \bf H}_E \right)^{\frac{1}{2}}\right)}
\end{equation}
This is due to the fact that the positive semidefinite constraint $\left(\left({ \bf H}_M^\dag { \bf H}_M \right)^{-1} - \nu\left({ \bf H}_E^\dag { \bf H}_E \right)^{-1}\right) \succeq {\bf 0}$
is equivalent to the constraint $\nu \leq \lambda_{min}\left(\left({ \bf H}_E^\dag { \bf H}_E \right)^{\frac{1}{2}} \left({ \bf H}_M^\dag { \bf H}_M \right)^{-1} \left({ \bf H}_E^\dag { \bf H}_E \right)^{\frac{1}{2}}\right)$, where
$\lambda_{min}\left({\bf M}\right)$ denotes the minimum eigenvalue of the positive definite
matrix ${\bf M}$. The solution to the optimization problem \eqref{dual_final1} -- \eqref{dual_final2} can be computed in a
straightforward manner using, for example, the bisection method~\cite{BurFai04}, which represents a much simpler procedure than any method that solves the non-linear equation in \eqref{nu}.

The optimal values of $\mu$ in \eqref{mu} and $\nu$, which corresponds to the solution of \eqref{dual_final1} subject to \eqref{dual_final2} then define the optimal transmit
filter. In turn, the optimal transmit filter defines the ZF receive filters through \eqref{Rx_m} and \eqref{Rx_e}.

\section {Optimal Linear Receive Filter at the Eavesdropper}
\label{wiezf_filters}

We now consider the scenario where the legitimate receiver uses a ZF filter, whilst the eavesdropper receiver uses the optimal linear Wiener filter. This corresponds to a generalization of the previous scenario where both the receivers are restricted to obey ZF constraints.

\subsection{Optimal Linear Receive Filter Design}
\label{rx_filters_wiener}

Let us consider the design of the eavesdropper optimal linear receive filter. Eve now uses the receive filter that, for any fixed transmit filter ${ \bf H}_T$, minimizes:
\begin{align}\label{MSE_Ea}
&{ {\sf MSE_{E}}  = \mathcal{E} \left[\|{\bf X} - {{\bf H}_R}_E
{\bf Y}_E\|^2\right]}
\end{align}
This corresponds to the Wiener filter given by (see e.g. \cite{Kay93}):
\begin{equation}
\label{rx_filter_eavesdropper} 
{ { \bf H}_{RE}^* = {\bf H}_T^\dagger{ \bf H}_E^\dagger\left({ \bf I} + { \bf H}_E{ \bf H}_T{ \bf H}_T^\dagger{ \bf H}_E^\dagger\right)^{-1}}
\end{equation}
In turn, the MSE in the eavesdropper channel, upon substituting \eqref{rx_filter_eavesdropper} in \eqref{MSE_Ea}, is given by:
\begin{equation} \label{MSE_E_wie}
{ {\sf MSE_{E}} ={\sf tr} \left\{\left({\bf I} + {\bf H}_E^\dag
{\bf H}_E {\bf H}_T {\bf H}_T^\dag\right)^{-1}\right\}}
\end{equation}

Note that the expressions for the legitimate receive filter and for the MSE in the the main channel are already given in \eqref{Rx_m} and \eqref{MSE_M}.

\subsection{Optimal Transmit Filters}
\label{tx_filters_wiezf}

We now consider the design of the optimal linear transmit filter. 
This, in view of \eqref{MSE_M} and \eqref{MSE_E_wie}, corresponds to the solution of the optimization problem given by:   
\begin{equation}\label{opt_1}
{ \min_{{\bf H}_T}~~{\sf tr}\left\{ \left({ \bf H}_T^\dag { \bf H}_M^\dag { \bf H}_M{ \bf H}_T\right)^{-1}\right\}}
\end{equation}
subject to the secrecy constraint:
\begin{equation}\label{opt_2}
{ {\sf tr} \left\{\left({\bf I} + {\bf H}_E^\dag {\bf H}_E {\bf
H}_T {\bf H}_T^\dag\right)^{-1}\right\} \geq \gamma}
\end{equation}
and to the power constraint:
\begin{equation}\label{opt_3}
{ {\sf tr}\left\{ { \bf H}_T { \bf H}_T^\dag \right\} \leq P_{avg}}
\end{equation}
with ${ \bf H}_T { \bf H}_T^\dag \succ {\bf 0}$. Note that -- due to the channel knowledge assumptions -- the legitimate transmitter, the legitimate receiver and the eavesdropper can also all set up this optimization problem to compute the transmit filter and receive filters via \eqref{Rx_m} and \eqref{rx_filter_eavesdropper}.

It is also possible to reduce this optimization problem to a standard convex optimization problem, by adopting the change of variables ${ \bf Z} = \left({ \bf H}_T { \bf H}_T^\dag \right)^{-1}$ together with the Woodbury matrix identity~\cite{Golub96}. Thus, the optimization problem reduces to:
\begin{equation}
{ \min_{{\bf Z}}~~{\sf tr}\left\{ \left({ \bf H}_M^\dag { \bf H}_M \right)^{-1} { \bf Z} \right\}}
\end{equation}
subject to the constraints:
\begin{equation}
{ {\sf tr}\left\{{\bf I}\right\} - {\sf tr}\left\{ \left({ \bf H}_E^\dag { \bf H}_E \right)\left( { \bf Z} + \left({ \bf H}_E^\dag { \bf H}_E \right)\right)^{-1}\right\} \geq \gamma}
\end{equation}
\begin{equation}
{ {\sf tr}\left\{ { \bf Z}^{-1} \right\} \leq P_{avg}}
\end{equation}
and ${\bf Z} \succ {\bf 0}$. The solution follows from the Karush-Kuhn-Tucker optimality conditions given by:
\begin{align}\label{kkt1}
& {  \left({ \bf H}_M^\dag { \bf H}_M \right)^{-1} - \nu\left[ \left({\bf Z} + \left({ \bf H}_E^\dag { \bf H}_E \right)\right)^{-1} \left({ \bf H}_E^\dag { \bf H}_E \right) \left({\bf Z} + \left({ \bf H}_E^\dag { \bf H}_E \right)\right)^{-1} \right] -~\mu{\bf Z}^{-2} = 0}
\end{align}
\begin{equation}
{  \nu \left\{{\sf tr}\left\{{\bf I}\right\} - {\sf tr}\left\{ \left({ \bf H}_E^\dag { \bf H}_E \right)\left( { \bf Z} + \left({ \bf H}_E^\dag { \bf H}_E \right)\right)^{-1}\right\} - \gamma \right\} = 0,~~~ \nu \geq 0 }
\end{equation}
\begin{equation}
{ \mu \left[ P_{avg} - {\sf tr}\left\{{\bf Z}^{-1}\right\} \right] = 0,~~~ \mu \geq 0}
\end{equation}
and $ {\bf Z} \succ {\bf 0}$, ${\sf tr}\left\{{\bf I}\right\} - {\sf tr}\left\{ \left({ \bf H}_E^\dag { \bf H}_E \right)\left( { \bf Z} + \left({ \bf H}_E^\dag { \bf H}_E \right)\right)^{-1}\right\} \geq \gamma$, ${\sf tr}\left\{ { \bf Z}^{-1} \right\} \leq P_{avg}$, where $\nu$ ans $\mu$ are the Lagrange multipliers associated with the secrecy and power constraints, respectively.

It is clear from the Karush-Kuhn-Tucker conditions above 
that there are three operational regimes: i) the scenario where the transmitter can use all the available power without violating the secrecy constraint, so that the secrecy constraint is not active ($\nu = 0$) and the power constraint is active ($\mu > 0$); ii) the scenario where both the secrecy and power constraints are active ($\nu > 0$ and $\mu > 0$); and iii) the scenario where the transmitter cannot use all the available power without violating the secrecy constraint, so that the secrecy constraint is active ($\nu > 0$) and the power constraint is inactive ($\mu = 0$). Note that this situation differs from the previous scenario (with ZF filters at both receivers) where it was possible to use all the power available without violating the secrecy constraint. The difference derives from the use of a more powerful receive filter by the eavesdropper. 

It is difficult to extract a characterization of the optimal filter design from the Karush-Kuhn-Tucker optimality conditions above in the general scenario, even though the problem is convex. Consequently, we concentrate on scenarios i) and iii) only.

\subsubsection{Power constraint active / secrecy constraint inactive}

This situation arises typically in a regime of low available power, due to the fact that the power, injected into the channel, is not enough to meet or violate the secrecy constraint. 

The following Theorem, which stems directly from the
Karush-Kuhn-Tucker optimality conditions above, defines the form of the optimal
transmit filter, in such a regime.
\vspace{0.20cm}
\begin{theorem}
\label{optimaltx_i}
Assume that the legitimate transmitter, the legitimate receiver and the eavesdropper know the exact channel matrices ${\bf H}_M$ and ${\bf H}_E$. Assume also that the legitimate receiver uses a ZF filter whereas the eavesdropper receiver uses the optimal linear Wiener filter. Then, an optimal transmit filter in the scenario where the power constraint is active whilst the secrecy constrain is inactive is, without loss of generality, given by:
\begin{align}\label{optfilt1}
&{ { \bf H}_T^*  = \alpha\left({ \bf H}_M^\dag { \bf H}_M \right)^{-\frac{1}{4}}}
\end{align}
where $\alpha = \sqrt{\frac{P_{avg}}{{\sf tr}\left\{\left({ \bf H}_M^\dag { \bf H}_M \right)^{-\frac{1}{2}}\right\}}}$.
\vspace{0.50cm}
\end{theorem}
Note that the right multiplication of the transmit filter in~\eqref{optfilt1} by any unitary matrix produces another optimal filter.

\begin{IEEEproof}
This Theorem follows from the Karush-Kuhn-Tucker conditions 
by using the fact that $\nu=0$, so that we can rewrite \eqref{kkt1} as follows:
\begin{align}
&{ \left({ \bf H}_M^\dag { \bf H}_M \right)^{-1} - \mu{\bf Z}^{-2} = 0}
\end{align}
\end{IEEEproof}

Note that, as expected, this solution corresponds to the solution embodied in Theorem \ref{optimaltx_zf}, when the secrecy constraint is inactive.

\subsubsection{Power constraint inactive / secrecy constraint active}

This is a situation that typically arises in a regime of high available power; in fact, the use of all the available power would immediately violate the secrecy constraint. 

The following Theorem, which also stems directly from the
Karush-Kuhn-Tucker optimality conditions, 
defines the form of the optimal
transmit filter, in such a regime.
In particular, we use the fact that there exists a non-singular $m \times m$ matrix ${\bf C}$ that diagonalizes both ${ \bf H}_M^\dag { \bf H}_M$ and ${ \bf H}_E^\dag { \bf H}_E$ simultaneously~\cite{Golub96}, i.e. ${\bf C}^{\dagger}{ \bf H}_E^\dag { \bf H}_E{\bf C} = {\bf \Lambda}_{E}$ and ${\bf C}^{\dagger}{ \bf H}_M^\dag { \bf H}_M{\bf C} = {\bf \Lambda}_{M}$,
where ${\bf \Lambda}_{M}$ and ${\bf \Lambda}_{E}$ are $m \times m$ positive definite diagonal matrices, with diagonal elements $\lambda_{M_i}$, $i=1, 2, \ldots, m$ and $\lambda_{E_i}$, $i=1,2, \ldots, m$, respectively.
\vspace{0.20cm}
\begin{theorem}
\label{optimaltx_ii}
Assume that the legitimate transmitter, the legitimate receiver and the eavesdropper know the exact channel matrices ${\bf H}_M$ and ${\bf H}_E$. Assume also that the legitimate receiver uses a ZF filter whereas the eavesdropper receiver uses the optimal linear Wiener filter. Then, an optimal transmit filter in the scenario where the power constraint is inactive whilst the secrecy constrain is active is, without loss of generality, given by:
\begin{align}\label{optfilt2}
&{ { \bf H}_T^* = {{\bf C}}\left(\alpha{\bf \Lambda}_{M}^{\frac{1}{2}}{\bf \Lambda}_{E}^{\frac{1}{2}} - {\bf \Lambda}_{E}\right)^{-\frac{1}{2}}}
\end{align}
where $\alpha = \frac{{\sf tr}\left\{{\bf \Lambda}_{E}^{\frac{1}{2}}{\bf \Lambda}_{M}^{-\frac{1}{2}}\right\}}{{\sf tr}\left\{{\bf I}\right\} - \gamma}$.
\vspace{0.50cm}
\end{theorem}
Note that the right multiplication of the transmit filter in~\eqref{optfilt2} by any unitary matrix produces another optimal filter.

\begin{IEEEproof}
This Theorem also follows from the Karush-Kuhn-Tucker conditions 
by using the fact that $\mu=0$, so that we can rewrite \eqref{kkt1} as follows:
\begin{align}
& {  \left({ \bf H}_M^\dag { \bf H}_M \right)^{-1} -~\nu\left[ \left({\bf Z} + \left({ \bf H}_E^\dag { \bf H}_E \right)\right)^{-1} \left({ \bf H}_E^\dag { \bf H}_E \right) \left({\bf Z} + \left({ \bf H}_E^\dag { \bf H}_E \right)\right)^{-1} \right] = 0}
\end{align}
or equivalently:
\begin{align}
{  {\bf \Lambda}_{M}^{-1} - \nu\left[ \left({\bf C}^{\dagger}{\bf Z}{\bf C} + {\bf \Lambda}_{E}\right)^{-1} {\bf \Lambda}_{E} \left({\bf C}^{\dagger}{\bf Z}{\bf C} + {\bf \Lambda}_{E}\right)^{-1} \right] = 0}
\end{align}
\end{IEEEproof}

\subsubsection{Interpretation}

It is interesting to contrast the operational principle of the optimal transmit filter design when the secrecy constraint is inactive (in Theorem \ref{optimaltx_i}) to that when the secrecy constraint is active (in Theorem \ref{optimaltx_ii}).

In the regime where the power constraint is active and the secrecy constraint is inactive, the optimal transmit filter decomposes the MIMO main channel into a set of parallel channels using an orthonormal transformation that does not affect the transmit power. The optimal transmit filter then weighs the individual subchannels, such that the power constraint is met with equality. The optimal weights depend only on the eigenvalues of the matrix ${ \bf H}_M^\dag { \bf H}_M$.

In the regime where the power constraint is inactive and the secrecy constraint is active, the optimal transmit filter decomposes simultaneously the MIMO main channel and the MIMO eavesdropper channel into a set of parallel channels using an in general non-orthonormal transformation. Note that, even though such a transformation may affect the transmit power, this is not a concern in this regime. The optimal transmit filter then weighs the individual subchannels further, such that the secrecy constraint is met with equality. Interestingly, the optimal weights now depend on the generalized eigenvalues of the matrices ${ \bf H}_M^\dag { \bf H}_M$ and ${ \bf H}_E^\dag { \bf H}_E$.

It is also interesting to contrast the transmit filter design when the eavesdropper employs a ZF filter (in Theorem \ref{optimaltx_zf}) to that when the eavesdropper employs a Wiener filter. In the ZF case, when the secrecy constraint is active, the transmit filter uses an orthonormal transformation to decompose an equivalent channel in view of the fact that the power constraint is always active. In the Wiener case, when the secrecy constraint is active, the transmit filter uses a non-singular matrix to decompose simultaneously both channels.

\subsection{A Note on the Validity of the Operational Regimes}
\label{regimes}

It is now relevant to establish conditions, which are a function of the system parameters, that identify the exact regions of validity of the operational regimes unveiled in the previous subsection.

\subsubsection{Power constraint active / secrecy constraint inactive}
To identify the validity of this regime we minimize the objective function in \eqref{opt_1}, subject to the power constraint in \eqref{opt_3} only. Note that this constitutes a relaxation of the original optimization problem so the solution of this new optimization problem can never lead to a worse MSE than the solution of the original problem. In turn, this solution is also a solution of the original optimization problem provided that it does not violate the secrecy constraint.

It is easy to show that this regime is valid if, for a fixed set of system parameters, $P_{avg}$, $\gamma$, ${ \bf H}_M$ and ${ \bf H}_E$, the following condition holds:
\begin{equation}\label{reg_i1}
{ 	{\sf tr}\left\{{\bf I}\right\} - {\sf tr}\left\{ { \bf H}_E^\dag { \bf H}_E\left[ \left({ \bf H}_T^* { \bf H}_T^{*\dag} \right)^{-1} + { \bf H}_E^\dag { \bf H}_E\right]^{-1}\right\} \geq \gamma}
\end{equation}
where ${\bf H}_T^*$ corresponds to the design embodied in Theorem \ref{optimaltx_i} given by:
\begin{equation}\label{reg_i2}
{ 	{ \bf H}_T^* = \sqrt{\frac{P_{avg}}{{\sf tr}\left\{\left({ \bf H}_M^\dag { \bf H}_M \right)^{-\frac{1}{2}}\right\}}} \left({ \bf H}_M^\dag { \bf H}_M \right)^{-\frac{1}{4}} }
\end{equation}

Note that \eqref{reg_i1} and \eqref{reg_i2} can be used to determine a threshold secrecy constraint, $\gamma_{max_{reg1}}$, below which we operate under this regime, or equivalently, a threshold power constraint, $P_{avg_{max R1}}$, below which we operate under this same regime. The threshold secrecy constraint is given by:
\begin{align}
& {  \gamma_{max_{reg1}} = {\sf tr}\left\{{\bf I}\right\} -~{\sf tr}\left\{ { \bf H}_E^\dag { \bf H}_E\left[ \frac{{\sf tr}\left\{\left({ \bf H}_M^\dag { \bf H}_M \right)^{-\frac{1}{2}}\right\}}{P_{avg}} \left({ \bf H}_M^\dag { \bf H}_M \right)^{\frac{1}{2}} + { \bf H}_E^\dag { \bf H}_E\right]^{-1}\right\}}
\end{align}

\subsubsection{Power constraint inactive / secrecy constraint active}

To identify the validity of this regime we now minimize the objective function in \eqref{opt_1}, subject to the secrecy constraint in \eqref{opt_2} only. This also constitutes a relaxation of the original optimization problem so the solution of this new optimization problem can never lead to a worse MSE than the solution of the original problem. Moreover, this solution is also a solution of the original optimization problem provided that it does not violate the power constraint.

It is also straightforward to show that this regime is valid if, for a fixed set of system parameters, $P_{avg}$, $\gamma$, ${ \bf H}_M$ and ${ \bf H}_E$, the following condition holds:
\begin{equation}\label{reg_ii1}
{ 	{\sf tr}\left\{ { \bf H}_T^* { \bf H}_T^{*\dag} \right\} \leq P_{avg} }
\end{equation}
where ${\bf H}_T^*$ corresponds to the design embodied in Theorem \ref{optimaltx_ii}, given by:
\begin{equation}\label{reg_ii2}
{  { \bf H}_T^* = {{\bf C}}\left( \frac{{\sf tr}\left\{{\bf \Lambda}_{E}^{\frac{1}{2}}{\bf \Lambda}_{M}^{-\frac{1}{2}}\right\}}{{\sf tr}\left\{{\bf I}\right\} - \gamma} {\bf \Lambda}_{M}^{\frac{1}{2}}{\bf \Lambda}_{E}^{\frac{1}{2}} - {\bf \Lambda}_{E}\right)^{-\frac{1}{2}} }
\end{equation}

Similarly to the previous case, \eqref{reg_ii1} and \eqref{reg_ii2} can be used to determine a threshold secrecy constraint, $\gamma_{min_{reg3}}$, above which we operate under this regime, or equivalently, a threshold power constraint, $P_{avg_{min R3}}$, above which we operate in the same regime. The threshold power constraint is given by:
\begin{equation}
{  P_{avg_{min R3}} = {\sf tr}\left\{ {{\bf C}}\left( \frac{{\sf tr}\left\{{\bf \Lambda}_{E}^{\frac{1}{2}}{\bf \Lambda}_{M}^{-\frac{1}{2}}\right\}}{{\sf tr}\left\{{\bf I}\right\} - \gamma} {\bf \Lambda}_{M}^{\frac{1}{2}}{\bf \Lambda}_{E}^{\frac{1}{2}} - {\bf \Lambda}_{E}\right)^{-1}{{\bf C}^{\dagger}} \right\}}
\end{equation}

\section{Generalizations}
\label{fading}

It is also of interest to generalize the filter design problem to scenarios that involve some degree of channel uncertainty. We consider two cases:

\begin{enumerate}

\item The legitimate receiver knows the exact state of the main channel and the statistics of the eavesdropper channel, the eavesdropper receiver knows the exact state of the eavesdropper channel and the statistics of the main channel, and the transmitter knows only the statistics of the main and eavesdropper channels;

\item The legitimate receiver knows the exact state of the main channel and the statistics of the eavesdropper channel, the eavesdropper receiver knows the exact state of the eavesdropper channel and the statistics of the main channel, and the transmitter knows the exact state of both channels.

\end{enumerate}

These scenarios arise naturally in the "secure" video broadcasting model depicted in Figure~\ref{fig:broadcast}, where both receivers -- even though they may have subscribed to different services -- are active users of the network: in case 1), it is assumed that the receivers convey information about the statistics of their own channels to the transmitter via a feedback path (this information is then relayed to the other receivers); in case 2), it is assumed that the receivers convey information about the exact state of their own channels to the transmitter also via a feedback path (this information is not relayed to the other receivers though)\footnote{\scriptsize Note that the transmitter may also be able to capture an estimate of the statistics of the channels or the state of the channels in time division duplex (TDD) environments.}. In addition, these scenarios can also be used to capture some of the uncertainty about the state of the eavesdropper channel leading to filter designs with considerable operational significance.

We also comment on more efficient mechanisms to use the available resources, due to the fact that some of the solutions unveiled earlier have demonstrated that the transmitter does not always use all the available power in order to meet the security constraints.

The ensuing formulations are based on the assumption that the so-called eavesdropper adopts a linear receiver. Once again, the implications of the use, by the eavesdropper, of a non-linear rather than linear estimator are also discussed in the Section \ref{results}.

\subsection{Scenario 1}

A possible formulation of the filter design problem when the receivers know the exact state of their own channels and the distribution of the other channels, whereas the transmitter knows only the distribution of the channels, is given by:
\begin{equation}\label{exp_a_i}
{ \min_{{\bf H}_T} {\sf \overline{MSE}_{M}} = \mathcal{E}_{{\bf H}_M, {\bf H}_E} \left\{{\sf MSE_M}\left({\bf H}_M, {\bf H}_E\right)\right\}}
\end{equation}
subject to the security constraint:
\begin{equation}\label{exp_b_i}
{\sf \overline{MSE}_{E}} = \mathcal{E}_{{\bf H}_M, {\bf H}_E} \left\{{\sf MSE_E}\left({\bf H}_M, {\bf H}_E\right)\right\} \geq \gamma
\end{equation}
and the total power constraint:
\begin{equation}\label{exp_c_i}
{\sf tr} \left\{{\bf H}_T {\bf H}_T^\dag\right\} \leq P_{avg}
\end{equation}
where ${\sf \overline{MSE}_{M}}$ is the expected value, with respect to ${\bf H}_M$ and ${\bf H}_E$, of the MSE in the main channel for fixed channel matrices ${\bf H}_M$ and ${\bf H}_E$, i.e. ${\sf MSE_M}\left({\bf H}_M, {\bf H}_E\right)$, and ${\sf \overline{MSE}_{E}}$ is the expected value, with respect to ${\bf H}_M$ and ${\bf H}_E$, of the MSE in the eavesdropper channel for fixed channel matrices ${\bf H}_M$ and ${\bf H}_E$, i.e. ${\sf MSE_E}\left({\bf H}_M, {\bf H}_E\right)$.

By assuming that the legitimate receiver uses a ZF filter and the eavesdropper uses either a ZF filter or a Wiener filter, then the optimization problem reduces to:

\begin{equation}\label{optfad1}
{ \min_{{\bf H}_T}~~\mathcal{E}_{{ \bf H}_M}\left\{ {\sf tr}\left\{ \left({ \bf H}_T^\dag { \bf H}_M^\dag { \bf H}_M{ \bf H}_T\right)^{-1}\right\} \right\}}
\end{equation}
subject to:
\begin{equation}\label{optfad3}
{ {\sf tr}\left\{ { \bf H}_T { \bf H}_T^\dag \right\} \leq P_{avg}}
\end{equation}
and:
\begin{equation}\label{optfad2}
{ \mathcal{E}_{{ \bf H}_E}\left\{ {\sf tr}\left\{ \left({ \bf H}_T^\dag { \bf H}_E^\dag { \bf H}_E{ \bf H}_T\right)^{-1}\right\} \right\} \geq \gamma}
\end{equation}
or:
\begin{equation}\label{optfad2_wie}
{ \mathcal{E}_{{ \bf H}_E}\left\{ {\sf tr}\left\{ \left({\bf I} + {\bf H}_E^\dag {\bf H}_E {\bf H}_T {\bf H}_T^\dag\right)^{-1}\right\} \right\} \geq \gamma}
\end{equation}
depending on whether it is assumed that the eavesdropper adopts a ZF or a Wiener filter, respectively.

The significance of this formulation relates to the fact that the legitimate transmitter, the legitimate receiver and the eavesdropper receiver all have the necessary information to set up this optimization problem in order to conceive the transmit filter and therefore the receive filters via \eqref{Rx_m} and \eqref{Rx_e} or \eqref{rx_filter_eavesdropper}, respectively. In addition, as long as the legitimate transmitter and the legitimate receiver agree to use this formulation to perform the legitimate transmit and receive filter designs, there is no incentive for the eavesdropper to adopt any other formulation beyond this one to design its own filter.

In particular, assume that the legitimate transmitter and the legitimate receiver adopt the formulation based on the use of a Wiener filter by the eavesdropper. If the eavesdropper adopted another linear filter, the average value of the MSE of the eavesdropper channel would still be above $\gamma$ in view of the optimality of the Wiener filter.

In contrast, assume that the legitimate transmitter and the legitimate receiver adopt the formulation based on the use of a ZF filter by the eavesdropper. In the regime of high available power, and once again if the eavesdropper used another linear filter, then the average value of the MSE of the eavesdropper channel would still be above $\gamma$ in view of the fact that the performance of a ZF filter approaches that of a Wiener filter in such a regime. In the regime of low available power, if the eavesdropper used a Wiener filter instead, then the average value of the eavesdropper MSE could be evidently below $\gamma$. This concern can be bypassed by operating at high enough available powers.

\subsection{Scenario 2}

A formulation of the filter design problem when the receivers know the exact state of their own channels and the distribution of the other channels, where as the transmitter knows the exact state of the channels, is given by:

\begin{equation}\label{exp_a_ii}
{ \min_{{\bf H}_T} {\sf MSE_M}\left({\bf H}_M, {\bf H}_E\right)}
\end{equation}
subject to the security constraint:
\begin{equation}\label{exp_b_ii}
{\sf \overline{MSE}_{E}} = \mathcal{E}_{{\bf H}_M, {\bf H}_E} \left\{{\sf MSE_E}\left({\bf H}_M, {\bf H}_E\right)\right\} \geq \gamma
\end{equation}
and the total power constraint:
\begin{equation}\label{exp_c_ii}
{\sf tr} \left\{{\bf H}_T {\bf H}_T^\dag\right\} \leq P_{avg}
\end{equation}

By assuming once again that the legitimate receiver uses a ZF filter and the eavesdropper uses either a ZF filter or a Wiener filter, then the optimization problem reduces to:
\begin{equation}\label{optfad1_ii}
{ \min_{{\bf H}_T}~~{\sf tr}\left\{ \left({ \bf H}_T^\dag { \bf H}_M^\dag { \bf H}_M{ \bf H}_T\right)^{-1}\right\} }
\end{equation}
subject to:
\begin{equation}\label{optfad3_ii}
{ {\sf tr}\left\{ { \bf H}_T { \bf H}_T^\dag \right\} \leq P_{avg}}
\end{equation}
and:
\begin{equation}\label{optfad2_ii}
{ \mathcal{E}_{{ \bf H}_E}\left\{ {\sf tr}\left\{ \left({ \bf H}_T^\dag { \bf H}_E^\dag { \bf H}_E{ \bf H}_T\right)^{-1}\right\} \right\} \geq \gamma}
\end{equation}
or:
\begin{equation}\label{optfad2_wie_ii}
{ \mathcal{E}_{{ \bf H}_E}\left\{ {\sf tr}\left\{ \left({\bf I} + {\bf H}_E^\dag {\bf H}_E {\bf H}_T {\bf H}_T^\dag\right)^{-1}\right\} \right\} \geq \gamma}
\end{equation}
depending on whether it is assumed that the eavesdropper adopts a ZF or a Wiener filter, respectively.

Note now that the legitimate transmitter and the legitimate receiver can also set up this optimization problem in order to determine the transmit filter and therefore the legitimate receive filter via \eqref{Rx_m}. In contrast, the eavesdropper -- in view of the absence of knowledge of the legitimate receiver channel -- cannot set up this optimization problem, so it is bound to use a mismatched filter. In view of the previous rationale, as long as the eavesdropper uses a linear filter and independently of whether the legitimate parties use the ZF or Wiener based formulation, we can thus argue that in the regime of high available power the average value of the eavesdropper MSE is always above $\gamma$ whereas in the regime of low available power the average value of the eavesdropper MSE can in principle be below $\gamma$, e.g. in the extremely unlikely event that the linear filter chosen (perhaps randomly) by the eavesdropper corresponds to the Wiener filter, but the legitimate parties assume that the eavesdropper uses a ZF rather than a Wiener filter in the design formulation.

Note also that this formulation does not explore the transmitter knowledge about the exact state of the eavesdropper channel per se. It is not clear whether or not such knowledge can be exploited in an operational meaningful way.

\subsection{Towards the solution of the new formulations}

These problems appear to be difficult to solve in general in view of the expectation operations in \eqref{exp_a_i} -- \eqref{exp_b_i} in scenario 1 and in \eqref{exp_b_ii} in scenario 2. However, it is possible to conceive a solution for the formulations that are based on the use of a ZF filter by the eavesdropper.

By adopting the change of variables ${ \bf Z} = \left({ \bf H}_T { \bf H}_T^\dag \right)^{-1}$ the optimization problem in \eqref{exp_a_i}, \eqref{exp_b_i} and \eqref{exp_c_i} reduces to:
\begin{equation}
{ \min_{{\bf Z}}~~{\sf tr}\left\{ \mathcal{E}_{{ \bf H}_M}\left\{\left({ \bf H}_M^\dag { \bf H}_M \right)^{-1} \right\} { \bf Z} \right\}}
\end{equation}
subject to:
\begin{equation}
\sf{ tr}\left\{ \mathcal{E}_{{ \bf H}_E}\left\{\left({ \bf H}_E^\dag { \bf H}_E \right)^{-1}\right\} { \bf Z} \right\} \geq \gamma
\end{equation}
and:
\begin{equation}
{\sf tr}\left\{ { \bf Z}^{-1} \right\} \leq P_{avg}
\end{equation}
and ${ \bf H}_T { \bf H}_T^\dag \succ {\bf 0}$, whereas the optimization problem in \eqref{exp_a_ii}, \eqref{exp_b_ii} and \eqref{exp_c_ii} reduces to:
\begin{equation}
{ \min_{{\bf Z}}~~{\sf tr}\left\{ \left({ \bf H}_M^\dag { \bf H}_M \right)^{-1} { \bf Z} \right\}}
\end{equation}
subject to:
\begin{equation}
\sf{ tr}\left\{ \mathcal{E}_{{ \bf H}_E}\left\{\left({ \bf H}_E^\dag { \bf H}_E \right)^{-1}\right\} { \bf Z} \right\} \geq \gamma
\end{equation}
and:
\begin{equation}
{\sf tr}\left\{ { \bf Z}^{-1} \right\} \leq P_{avg}
\end{equation}

The availability, when ${ \bf H}_M$ is such that its $n_M$ rows are independent $\mathcal{CN}\left(0,{\bf \Sigma}_M\right)$ circularly symmetric complex Gaussian random vectors and when ${ \bf H}_E$ is such that its $n_E$ rows are also independent $\mathcal{CN}\left(0,{\bf \Sigma}_E\right)$ circularly symmetric complex Gaussian random vectors, of closed form expressions for $\mathcal{E}\left\{\left({ \bf H}_M^\dag { \bf H}_M \right)^{-1} \right\}$ and $\mathcal{E}\left\{\left({ \bf H}_E^\dag { \bf H}_E \right)^{-1}\right\}$, which are given by~\cite{Mui82}:
\begin{equation}
{  \mathcal{E}_{{ \bf H}_M}\left\{\left({ \bf H}_M^\dag { \bf H}_M \right)^{-1}\right\} = \frac{1}{n_M-m-1}{\bf \Sigma}_M^{-1}, ~~~ \text{for}~ n_M-m-1 > 0}
\end{equation}
and 
\begin{equation}
{  \mathcal{E}_{{ \bf H}_E}\left\{\left({ \bf H}_E^\dag { \bf H}_E \right)^{-1}\right\} = \frac{1}{n_E-m-1}{\bf \Sigma}_E^{-1}, ~~~ \text{for}~ n_E-m-1 > 0}
\end{equation}
enable us to solve the optimization problem using the previous techniques~\cite{Mui82}. 

The availability of closed for expressions for $\mathcal{E}_{{ \bf H}_M}\left\{\left({ \bf H}_M^\dag { \bf H}_M \right)^{-1} \right\}$ and $\mathcal{E}_{{ \bf H}_E}\left\{\left({ \bf H}_E^\dag { \bf H}_E \right)^{-1}\right\}$ when ${ \bf H}_M$ and ${ \bf H}_E$ follow more general distributions would allow us to solve the optimization problem in other scenarios too.

\subsection{A discussion about effective use of resources}

Another relevant aspect relates to the fact that some of the filter designs are such that the transmitter does not use the entire available power budget in order to meet the secrecy constraint (see Section \ref{wiezf_filters}). One could thus argue that there is not an effective use of the available resources.

There are various possible generalizations to address this issue:

\subsubsection{Enter artificial noise} Artificial noise is an effective approach to provide some degree of distortion at the eavesdropper (\cite{GoeNeg08},~\cite{PeiWeiWongWang2012},~\cite{MukSwi2011} and~\cite{MukSwi09}), so it is interesting to reflect whether it might be possible to integrate elements of the filter design approach with elements of the artificial noise paradigm whereby the fraction of the unused power is also explored to further jam the eavesdropper. 

In general, it is not possible to integrate directly the artificial noise approach with our filter design approach because the transmitter does not signal over the null space of the main channel. 

However, it is possible to conceive more elaborate scenarios that involve the use of an additional
friendly jammer that shares the available power budget with the transmitter. This jammer is also constrained to convey artificial noise over the null space of the MIMO channel that links the jammer to the legitimate
receiver.

The action of the jammer -- which adds additional noise to the eavesdropper channel -- translates into a
new eavesdropper channel between the transmitter and the eavesdropper receiver incorporating the effect
of the artificial noise, that replaces the original eavesdropper channel. Therefore, one can pose immediately
an optimization problem akin to the previous filter design with secrecy constraints optimization problems
that -- in addition to involve the design of the transmit filter -- also involves the determination of the fraction
of power to be used by the legitimate transmitter and the fraction of power to be used by the friendly
jammer subject to the available power budget. The determination of the solution of this optimization
problem entails the extra level of complexity associated with how to share the power budget though.

\subsubsection{Enter the time and frequency dimension} Another approach that points towards a more efficient use of the resource relates to scenarios where one leverages the variability of the channel in the time domain (as in MIMO wireless channels) or in the frequency domain (as in MIMO-OFDM channels) in conjunction with available power constraints that operate along the multiple dimensions, i.e. long-term -- rather than short-term -- power constraints (e.g.~\cite{CaiTarBig99},~\cite{PraVar05} and~\cite{KhoLan12}). As an example, by assuming that all the parties know the state of the various time and/or frequency channels, it is possible to put forth the optimization problems:
\begin{equation}
{ \min_{{\bf H}_T(i),~~i=1,\cdots,n} ~~ \frac{1}{n} \sum_{i=1}^{n} {\sf MSE_M}\left({\bf H}_M(i), {\bf H}_E(i)\right)}
\end{equation}
subject to:
\begin{equation}
{\frac{1}{n} \sum_{i=1}^{n} {\sf MSE_E}\left({\bf H}_M(i), {\bf H}_E(i)\right)} \geq \gamma
\end{equation}
and:
\begin{equation}
{\sf tr} \left\{{\bf H}_T(i) {\bf H}_T(i)^\dag\right\} \leq P_{avg},~~i=1,\cdots,n
\end{equation}
assuming a short-term power constraint, or:
\begin{equation}
\frac{1}{n} \sum_{i=1}^{n} {\sf tr} \left\{{\bf H}_T(i) {\bf H}_T(i)^\dag\right\} \leq P_{avg}
\end{equation}
assuming a long-term power constraint, where ${\bf H}_T(i)$ is the transmit filter at time/frequency $i$ and ${\bf H}_M(i)$ and ${\bf
H}_E(i)$ contain the gains from each main and eavesdropper channel input to each main and eavesdropper channel output, respectively, at time/frequency $i$.

The use of the long-term power constraint -- instead of the short-term one -- now offers the means to distribute the available power more efficiently over the time or frequency dimensions in order to obtain a better performance. Note that the short-term power constraint filter design problem can leverage the previous techniques (see Sections \ref{ZF_filters} and \ref{wiezf_filters}); on the other hand, the long-term power constraint problem may require more sophisticated techniques.

\section{Numerical Results}
\label{results}

We now present a set of numerical results in order to provide further insight into the problem of filter design with secrecy constraints. In particular, we present the performance of the filter designs in the presence of perfect and imperfect channel knowledge, as well as in the presence of eavesdroppers that adopt non-linear rather than linear estimation. We also present the impact of the filter designs on other relevant metrics, that include the error probability and achievable secrecy rates. We consider for simplicity a $2 \times 2$ MIMO Gaussian wiretap channel where the main channel and the eavesdropper channel matrices are, respectively, given by:
\begin{equation}
{{\bf H}_M= \left[\begin{matrix} 4 & {-1} \\ 1 & 2 \end{matrix}\right]} \qquad , \qquad {{\bf H}_E= \left[\begin{matrix} 2 & {-1} \\ 1 & 1 \end{matrix}\right]} \nonumber
\end{equation}
This constitutes a degraded scenario because $\mathbf{H}_M^{\dagger}\mathbf{H}_M \succ \mathbf{H}_E^{\dagger}\mathbf{H}_E$, therefore, in general the MSE in the eavesdropper channel will be higher than the one in the main channel.

\subsection{Performance of the Filter Designs in the Presence of Perfect Channel Knowledge}

\begin{figure}[!htbp]
    \begin{center}
        \mbox{\includegraphics[width=5.70in , height=3in]{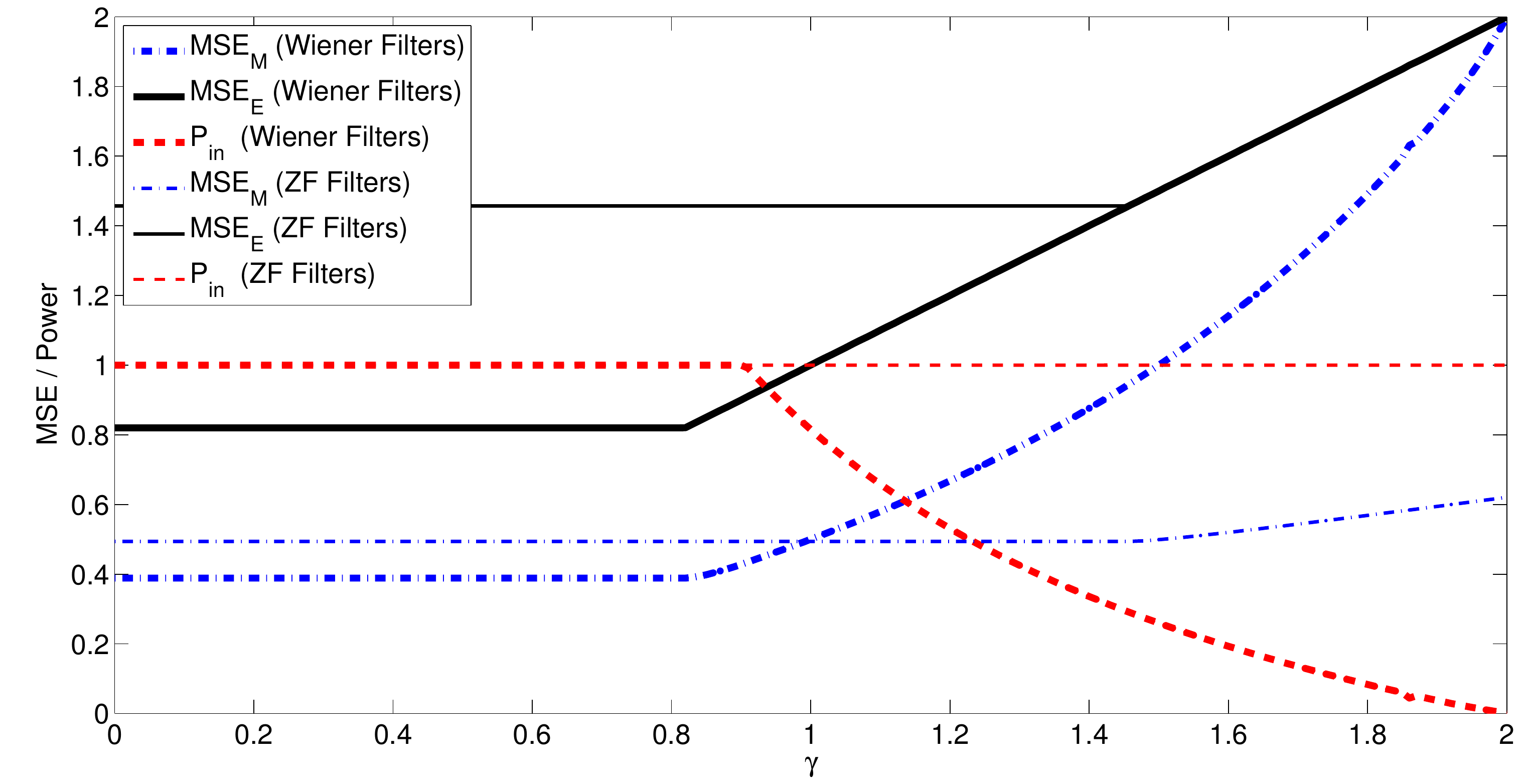}}
    \vspace{-0.5cm}
    \caption{Main and eavesdropper channel MSEs \emph{vs}. secrecy constraint and input power \emph{vs}. secrecy constraint, for the optimal transmit filter design and either ZF filters at both receivers or Wiener filters at both the receivers  ($P_{avg} = 1$).}
    
    \label{fig:ZFvsWie}
    \end{center}
    \label{ZFvsWie}
\end{figure}

We first consider the scenario where the channels are known perfectly by all the nodes -- as assumed in Theorems \ref{optimaltx_zf}, \ref{optimaltx_i} and \ref{optimaltx_ii} -- in order to test the performance of our designs. Figure \ref{fig:ZFvsWie} depicts the MSEs in the main and in the eavesdropper channels and the input power to the channels \emph{vs}. the secrecy constraint for $P_{avg} = 1$ when ZF filters are used at both the receivers. The solution clearly depicts the two operational regimes unveiled in Theorem \ref{optimaltx_zf}: i) the regime where the power constraint is active but the security constraint is inactive (for smaller values of $\gamma$); and ii) the regime where the power and security constraints are active and met with equality (for larger values of $\gamma$). Figure \ref{fig:ZFvsWie} also depicts the MSEs in the main and in the eavesdropper channels and the input power to the channels vs. the secrecy constraint for $P_{avg} = 1$ when the optimal linear Wiener filters are used at both receivers, in order to provide further insight.\footnote{\scriptsize To the best of our knowledge, the problem of filter design with secrecy constraints when Wiener filters are used at both receivers is not a convex in general. Therefore, an approximate solution has been determined through numerical
methods.} Surprisingly, in the relevant regime of large $\gamma$, the use of ZF filters rather than Wiener filters leads to a better MSE in the main channel without the violation of the security constraint. This is due to the fact that -- via the use of ZF filters in \emph{lieu} of the Wiener ones -- the transmitter can use all of the available power in such a scenario, in order to drive the MSE to a lower value.



\begin{figure}[!htbp]
    \begin{center}
        \mbox{\includegraphics[width=5.70in , height=3in]{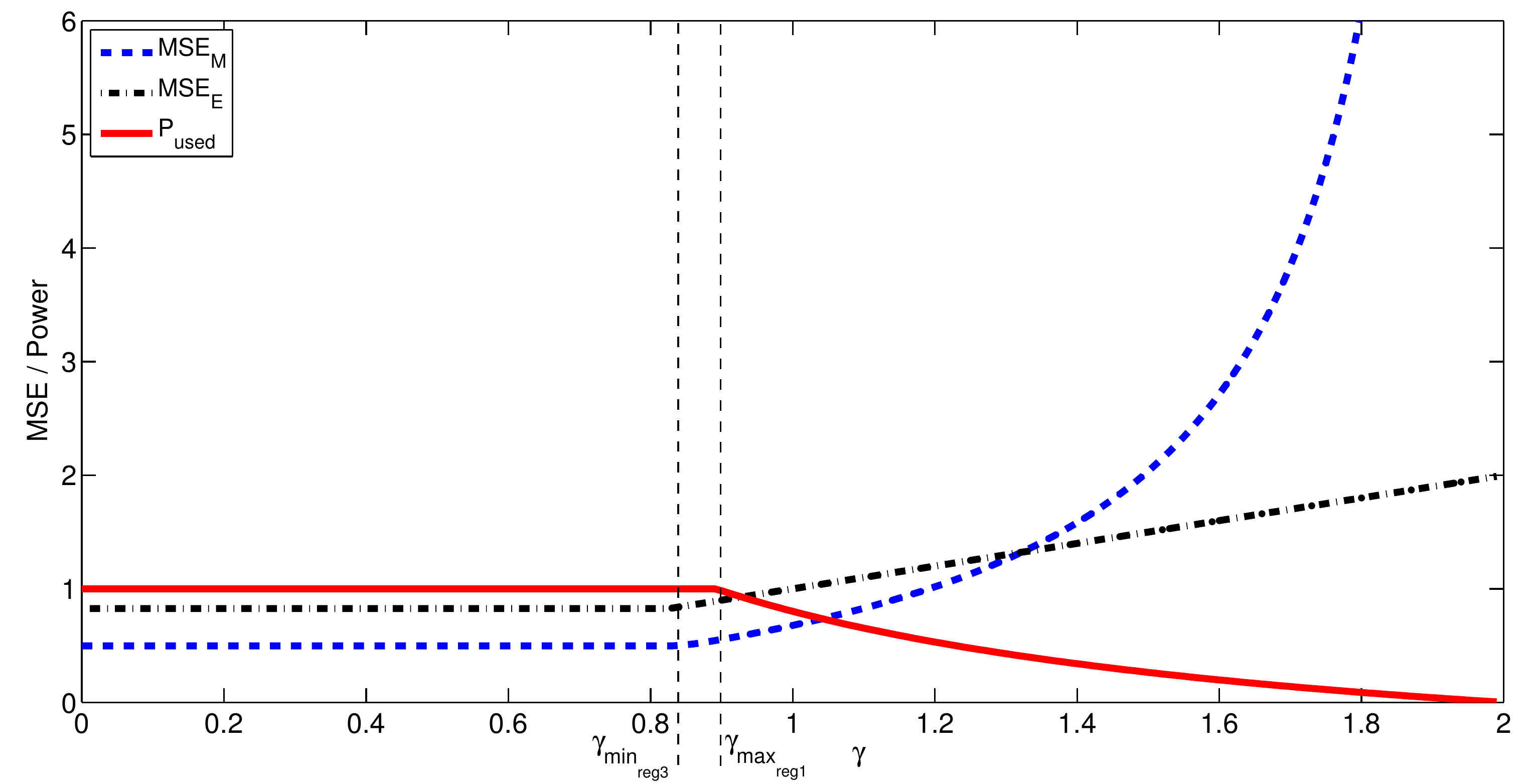}}
        \vspace{-0.5cm}
   \caption{Main and eavesdropper channel MSEs \emph{vs}. secrecy constraint and input power \emph{vs}. secrecy constraint, for the optimal transmit filter design with a ZF filter at the legitimate receiver and a Wiener filter at the eavesdropper receiver ($P_{avg} = 1$).}
    \label{fig:ZFvsWIE_gamma}
    \end{center}
    \label{ZFvsWIE_gamma}
\end{figure}

Figure \ref{fig:ZFvsWIE_gamma} now shows the values of the MSEs in the main and in the eavesdropper channels and the injected power into the channels \emph{vs}. the secrecy constraint for $P_{avg}=1$, when the eavesdropper uses the optimal linear filter instead. The solution exhibits the three operational regimes characterized in Section \ref{tx_filters_wiezf}. Below $\gamma_{max_{reg1}}$, the optimal transmit filter, which is given by Theorem \ref{optimaltx_i}, minimizes the MSE in the main channel subject to the power constraint only. We can indeed verify that the available power is not sufficient to meet or violate the secrecy constraint. In-between $\gamma_{max_{reg1}}$ and  $\gamma_{min_{reg3}}$,the transmit filter\footnote{\scriptsize The solution in this regime, which has not been derived, was obtained through numerical methods.} minimizes the MSE in the main channel while meeting the power and the secrecy constraint with equality. Above  $\gamma_{min_{reg3}}$, the optimal transmit filter, which is given by Theorem \ref{optimaltx_ii}, minimizes the MSE in the main channel subject to the secrecy constraint only. Note that it is not possible to use all the available power, otherwise the secrecy constraint would be violated. This power restriction results in a much higher MSE in the main channel than in the eavesdropper channel for large values of $\gamma$ because as the injected power tends to zero the MSE that results from the ZF receiver grows very rapidly.

Finally, in view of the fact that we have motivated the filter design problem with secrecy constraints problems in scenarios where a provider seeks to guarantee that users that have subscribed to a service have a reasonable quality of service, whereas users that did not do not experience such quality of service, it is relevant to understand whether or not there are circumstances where the MSE in the main channel can in fact be higher than the MSE in the eavesdropper channel.

In the presence of channel degradedness the main channel MSE can be higher than the eavesdropper channel MSE for low available power $P_{avg}$ for a fixed target $\gamma$ when the legitimate receiver uses a ZF filter and the eavesdropper receiver uses the Wiener filter. However, with the increase in the available power the performance of the ZF filter approaches that of the Wiener filter, so that -- in view of channel degradedness - the main channel MSE eventually becomes lower than the eavesdropper channel MSE.

\begin{figure}[!htbp]
    \begin{center}
        \mbox{\includegraphics[width=5.70in , height=3in]{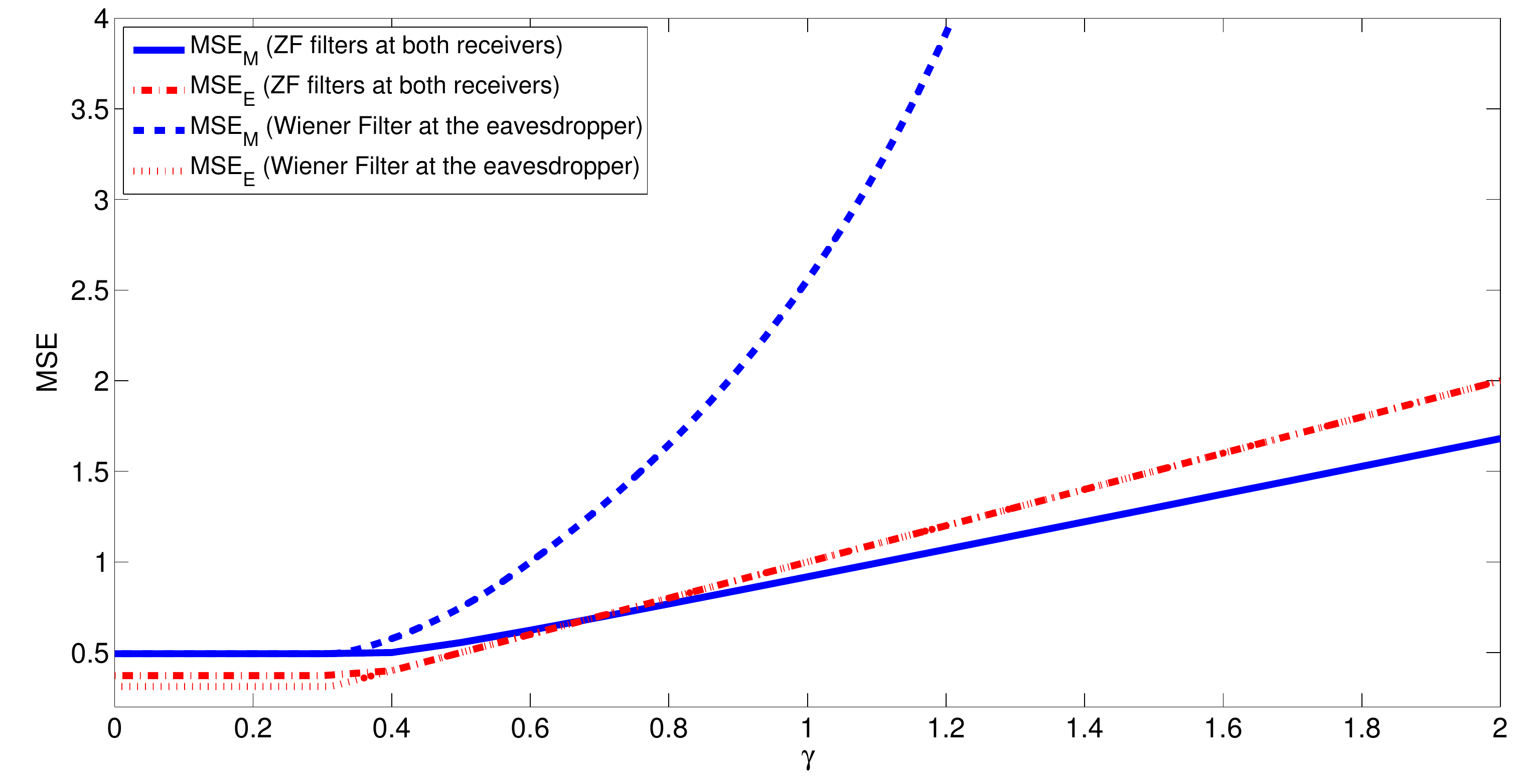}}
    \vspace{-0.5cm}
    \caption{Main and eavesdropper channel MSEs \emph{vs}. secrecy constraint, for the optimal transmit filter design with ZF filters at both receivers and Wiener filters at the eavesdropper receiver, in a non-degraded scenario  ($P_{avg} = 1$).}
    
    \label{fig:non_deg}
    \end{center}
    \label{non_deg}
\end{figure}

In contrast, in the absence of channel degradedness the main channel MSE can be higher than the eavesdropper channel MSE when both the legitimate receiver and the eavesdropper receiver use ZF filters or when the legitimate receiver uses a ZF filter and the eavesdropper receiver uses the Wiener one. This aspect is highlighted for a scenario where $H_M = \left[ \begin{smallmatrix} 4 & {-1}\\ 1 & 2 \end{smallmatrix} \right]$ and $H_E =\left[ \begin{smallmatrix} {3.5} & {-1}\\ 1 & 3 \end{smallmatrix} \right]$ in Figure \ref{fig:non_deg} -- note that MSE of
the eavesdropper obeys the secrecy constraint though.

However, with the emergence of MIMO-OFDM systems in a variety of wireless standards, it is possible conceive approaches that bypass the absence of degradedness. For example, one can in principle select sets of sub-carriers whose MIMO channels obey the degradedness property in order to assure that the MSE in the main channel is significantly lower than the MSE in the eavesdropper channel.

\subsection{Performance of the Filter Designs in the Presence of Imperfect Channel Knowledge}

We now consider the scenario where the channels are only known imperfectly by the nodes in order to test the robustness of the designs embodied in Theorems \ref{optimaltx_zf}, \ref{optimaltx_i} and \ref{optimaltx_ii}. In particular, we assume that the nodes have only access to an estimate of the main channel ${\tilde{\bf H}}_M = {\bf H}_M + {\bf \Phi}_M$, where ${\bf H}_M$ represents the true main channel matrix and ${\bf \Phi}_M$ models the main channel estimation error (with i.i.d. elements that follow a Gaussian distribution with mean zero and variance $\sigma_M^2$), as well as access to an estimate of the eavesdropper channel ${\tilde{\bf H}}_E = {\bf H}_E + {\bf \Phi}_E$, where ${\bf H}_E$ represents the true eavesdropper channel matrix and ${\bf \Phi}_E$ models the eavesdropper channel estimation error (also with i.i.d. elements that follow a Gaussian distribution with mean zero and variance $\sigma_E^2$). We also assume, for simplicity, that all the nodes have access to exactly the same estimates of the main and eavesdropper channel. The transmit and receive filters are designed based on the estimate of the channels rather than the true channels, via Theorems \ref{optimaltx_zf}, \ref{optimaltx_i} and \ref{optimaltx_ii}.

\begin{figure}[!htbp]
    \begin{center}
        \mbox{\includegraphics[width=5.70in , height=3in]{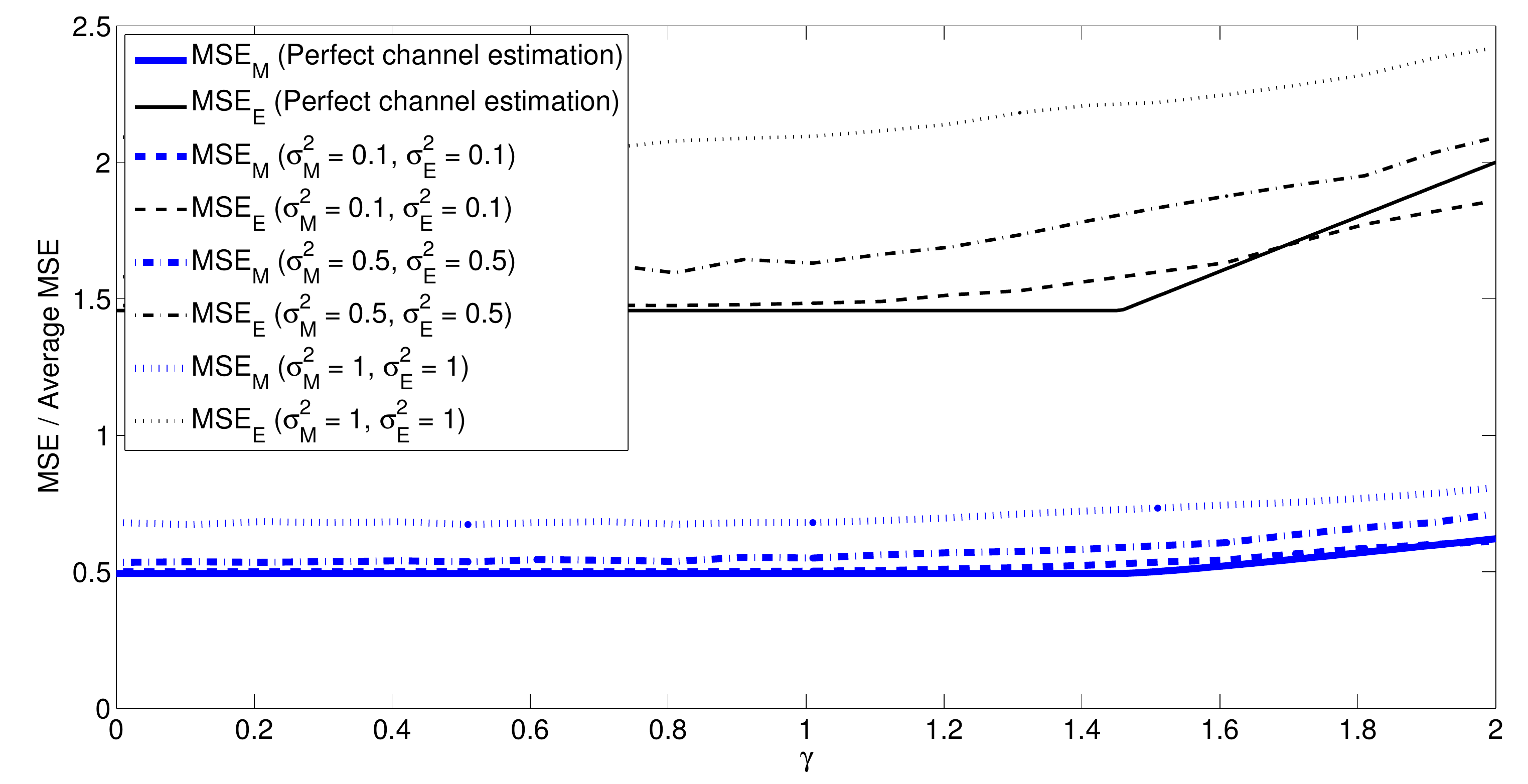}}
        \vspace{-0.5cm}
   \caption{Main and eavesdropper channel average MSEs \emph{vs}. secrecy constraint, in the presence of channel error estimation, for the optimal transmit filter design with ZF filter at both receivers ($P_{avg} = 1$).}
    \label{fig:est_err_zf}
    \end{center}
    \label{est_err_zf}
\end{figure}

\begin{figure}[!htbp]
    \begin{center}
        \mbox{\includegraphics[width=5.70in , height=3in]{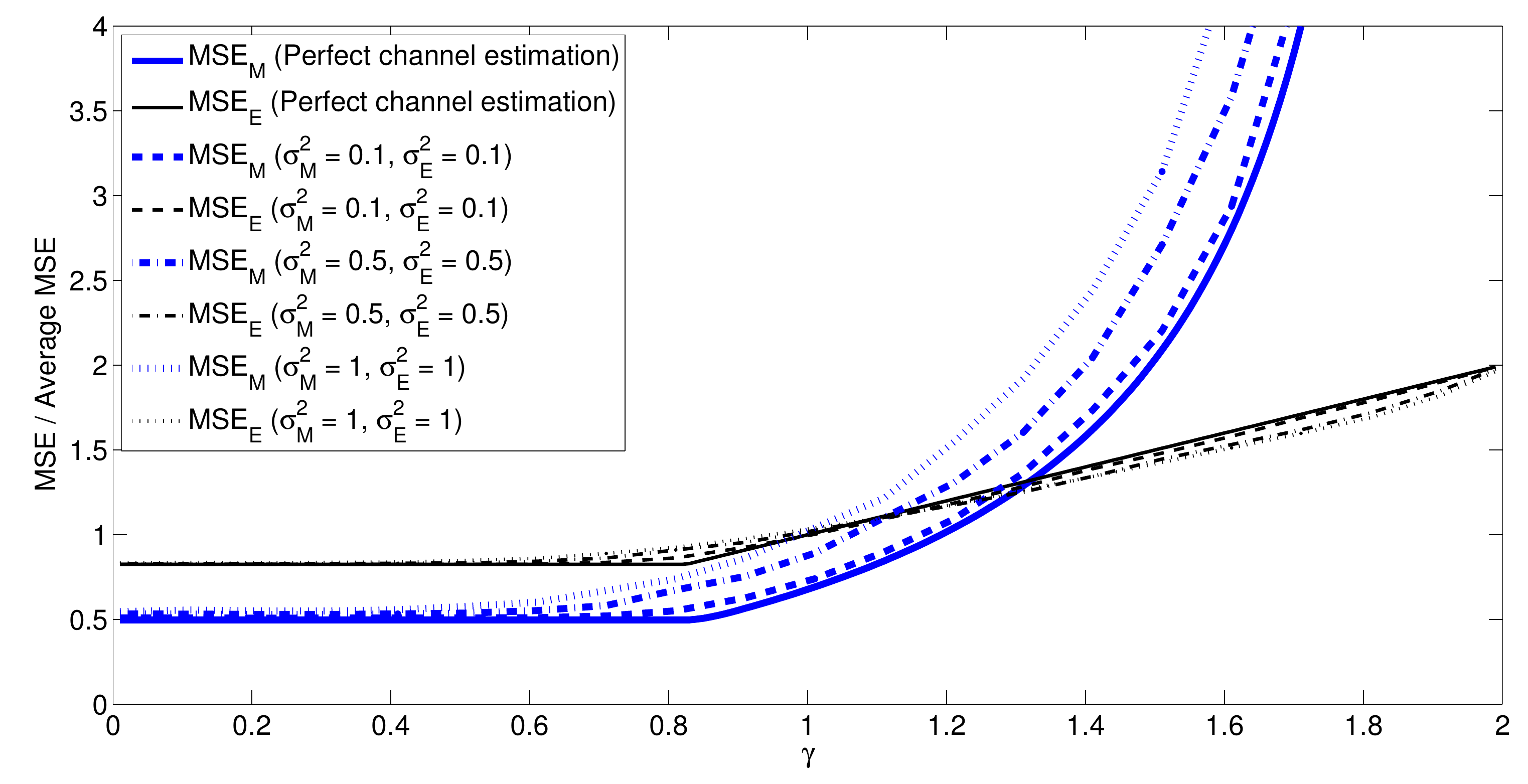}}
        \vspace{-0.5cm}
   \caption{Main and eavesdropper channel average MSEs \emph{vs}. secrecy constraint, in the presence of channel error estimation, for the optimal transmit filter design with a ZF filter at the legitimate receiver and a Wiener filter at the eavesdropper receiver ($P_{avg} = 1$).}
    \label{fig:est_err_wie}
    \end{center}
    \label{est_err_wie}
\end{figure}

Figures \ref{fig:est_err_zf} and \ref{fig:est_err_wie} depict the MSEs in the main and eavesdropper channels (averaged over 2000 realizations of the matrices that model the channel estimation errors) \emph{vs.} the secrecy constraint for $P_{avg = 1}$, for the scenario where the legitimate and eavesdropper receivers use ZF filters and the scenario where the legitimate receiver uses a ZF filter but the eavesdropper uses a Wiener filter, respectively.

We observe that channel modelling errors have an impact on the MSE of the main channel and -- of particular relevance -- on the MSE of the eavesdropper channel. The higher the deviation of the channel estimate from the true channel, which is modelled by the variances $\sigma_M^2$ and $\sigma_E^2$, the higher the deviation of the new MSEs from the original ones.

However, we also observe that the filter designs exhibit a certain degree of robustness. In the scenario where the eavesdropper uses the Wiener filter, the corresponding MSE appears to be reasonably robust to the channel modelling errors. In contrast, in the scenario where the eavesdropper uses a ZF filter, the corresponding MSE is more sensitive to the channel modelling errors. 

In general, for low to moderate channel estimation errors, the filter designs still guarantee that the secrecy constraint is not violated for a reasonable large set of $\gamma$.

\subsection{Linear vs. Nonlinear Estimation}

It is also relevant to consider the situation where the eavesdropper is not restricted to choose a linear filter. One could in principle argue that the eavesdropper (even if another user of a network as in Figure~\ref{fig:broadcast}) could use the optimal nonlinear receive filter, instead of the optimal linear one, to process the information in order to derive a lower MSE. This involves using a conditional mean estimator (CME), that delivers the estimate given by:
\begin{align}
\lefteqn{{\bf \hat{X}}_E = \mathcal{E}\left\{ {\bf X} \mid {\bf Y}_E \right\} =} \nonumber \\
&\frac{\int {\bf x}~~P_{\bf X}\left({\bf X}={\bf x}\right)~~P_{{\bf Y}_E \mid {\bf X}}\left({\bf Y}_E \mid {\bf X} = {\bf x}\right) d{\bf x}}{\int P_{\bf X}\left({\bf X}={\bf x}\right)~~P_{{\bf Y}_E \mid {\bf X}}\left({\bf Y}_E \mid {\bf X} = {\bf x}\right) d{\bf x}}
\end{align}
where $P_{\bf X}\left({\bf X}\right)$ is the probability density function of the input and $P_{{\bf Y}_E \mid {\bf X}}\left({\bf Y}_E \mid {\bf X}\right)$ is the conditional probability density function of the eavesdropper receive vector ${\bf Y}_E$ given the input vector ${\bf X}$.

\begin{figure}[!htbp]
    \begin{center}
        \mbox{\includegraphics[width=5.70in , height=3in]{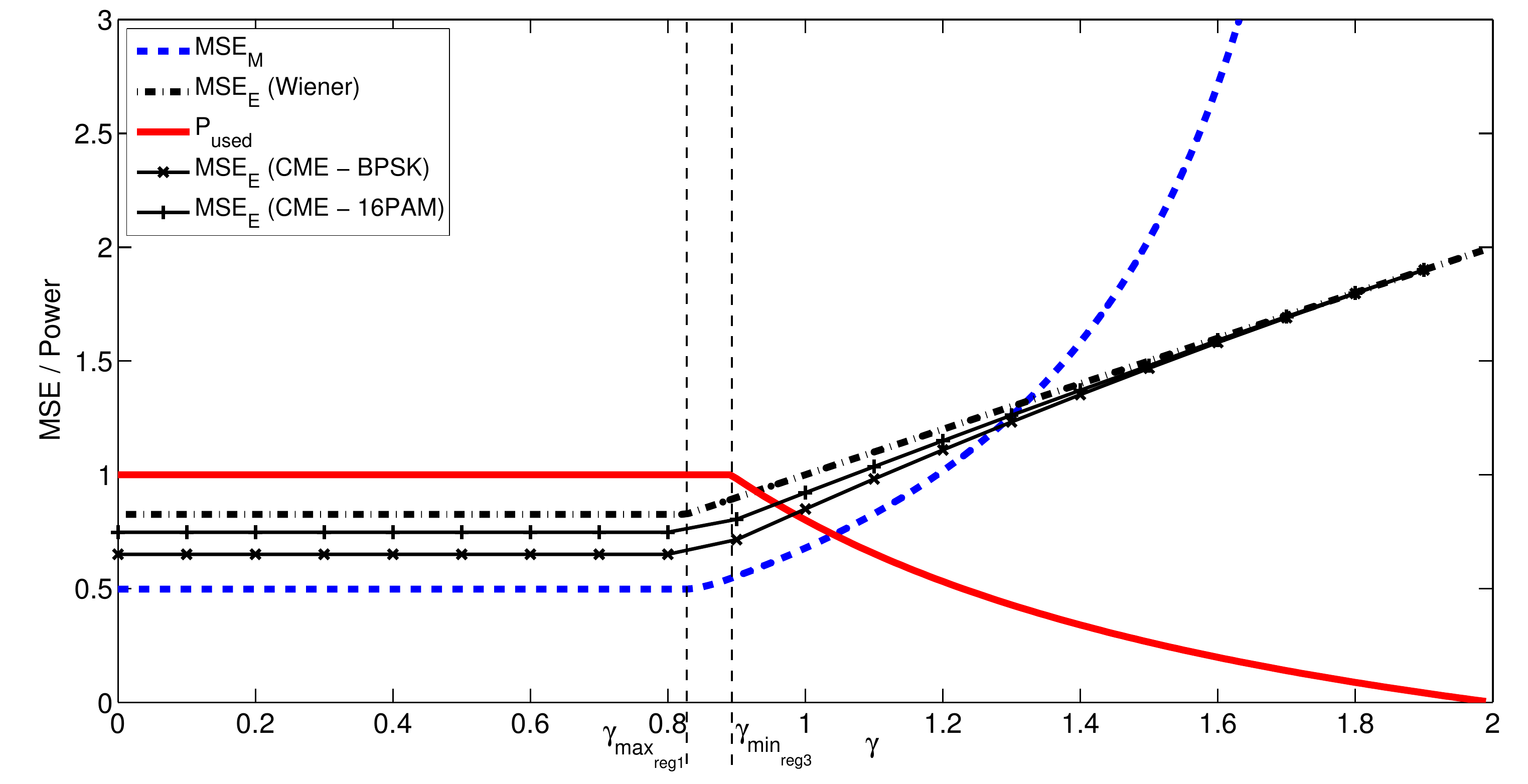}}
    \vspace{-0.5cm}
    \caption{Main and eavesdropper channel MSEs \emph{vs}. secrecy constraint and input power \emph{vs}. secrecy constraint, for the transmit filter design based on the use of a ZF filter at the legitimate receiver and a Wiener filter at the eavesdropper ($P_{avg} = 1$). ${\sf MSE_E (Wiener)}$ corresponds to the eavesdropper MSE associated with the linear Wiener filter. ${\sf MSE_E (CME - BPSK)}$ corresponds to the eavesdropper MSE associated with the CME for BPSK inputs. ${\sf MSE_E (CME - 16PAM)}$ corresponds to the eavesdropper MSE associated with the CME for 16PAM inputs.}
    
    \label{fig:CME1}
    \end{center}
    \label{CME1}
\end{figure}

We thus assess the performance penalty incurred by the use of a conditional mean estimator by the eavesdropper, but the transmitter designs its filter based on the assumption that the eavesdropper uses the optimal linear filter. We study scenarios where the elements of the input vector ${\bf X}$ are either BPSK or 16-PAM. Figure \ref{fig:CME1} shows the values of the MSEs in the main and in the eavesdropper channels and the injected power into the channels \emph{vs}. the secrecy constraint for $P_{avg}=1$. We can observe that designing the transmit filters based on the assumption that the eavesdropper is using an optimal linear receive filter results, as expected, in a lower eavesdropper MSE, when the input is not Gaussian (note that for Gaussian signals the conditional mean estimator is, in fact, linear). However, and interestingly, in regimes of greatest operational interest of large $\gamma$, the penalty that we pay by assuming that the eavesdropper uses an optimal linear filter rather than the optimal non-linear one vanishes, so that the eavesdropper does not have any real advantage in using the considerably more complex conditional mean estimator. This is due to the fact that the power injected in the channel approaches zero as the values of $\gamma$ increases, in order to meet the secrecy constraint.

\subsection{Impact of the Filter Designs on Other Metrics}

It is also of interest to assess the impact of the filter designs on other metrics of operational relevance, including the Bit Error Rate (BER) in the main and eavesdropper channels as well as achievable secrecy rates.

\begin{figure}[!htbp]
    \begin{center}
        \mbox{\includegraphics[width=5.70in , height=3in]{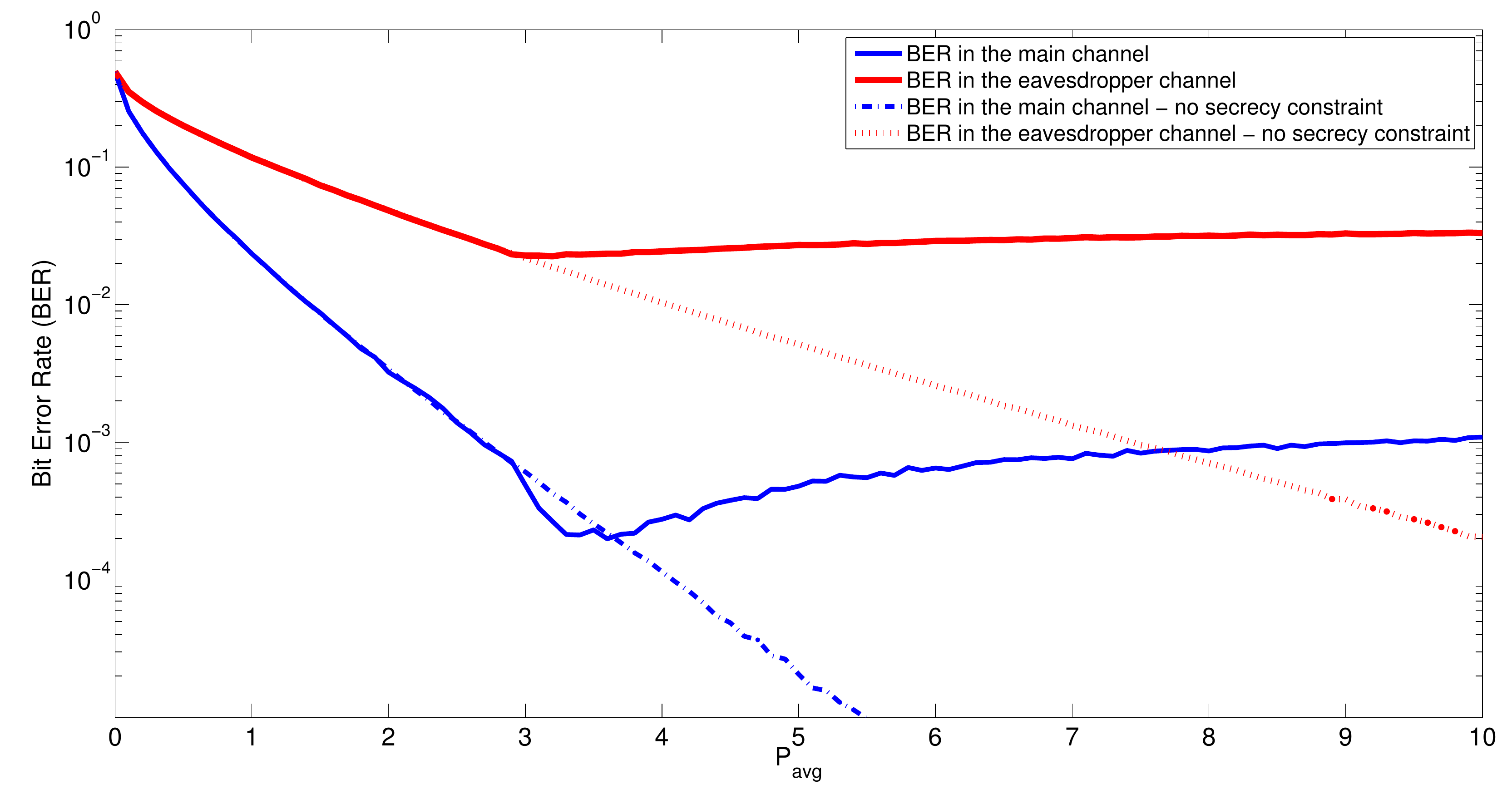}}
    \vspace{-0.5cm}
    \caption{Bit error rate \emph{vs}. available power for the scenario where both receivers use ZF filters ($\gamma = 0.5$).}
    
    \label{fig:BER1}
    \end{center}
    \label{BER1}
\end{figure}

\begin{figure}[!htbp]
    \begin{center}
        \mbox{\includegraphics[width=5.70in , height=3in]{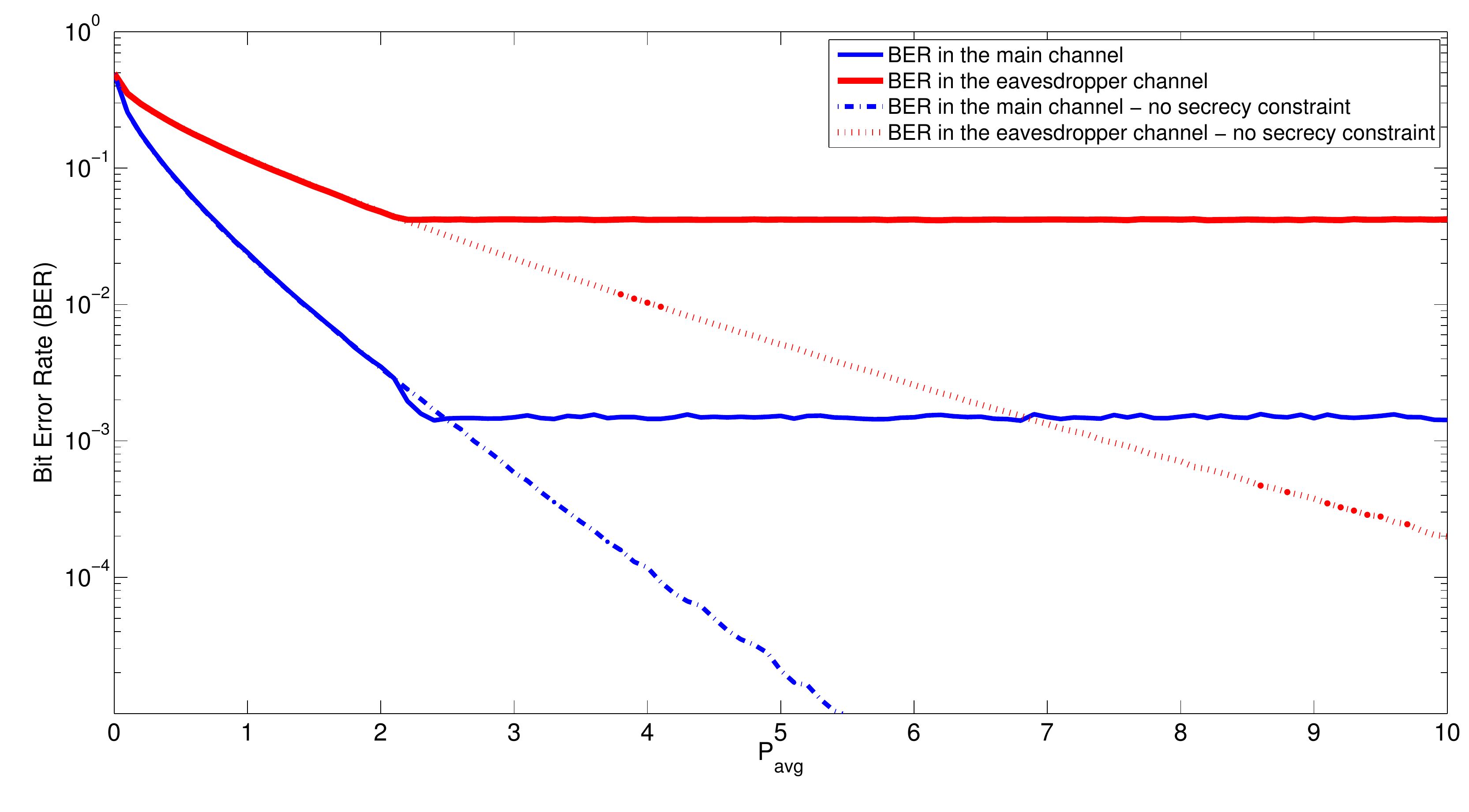}}
    \vspace{-0.5cm}
    \caption{Bit error rate \emph{vs}. available power for the scenario where the legitimate receiver uses a ZF filter and the eavesdropper receiver uses the optimal linear filter ($\gamma = 0.5$).}
    
    \label{fig:BER2}
    \end{center}
    \label{BER2}
\end{figure}

Figure \ref{fig:BER1} and \ref{fig:BER2} depict the Bit Error Rates (BER) of the main and the eavesdropper channels for the scenarios where i) ZF filters are used at both receivers and ii) a ZF receiver is used at the legitimate receiver and a Wiener filter is used at the eavesdropper receiver, respectively. These BER results are obtained through Monte Carlo simulations, assuming that the transmitter uses BPSK modulation and that the receiver uses a simple slicer to detect the information at the filters output.
We can observe that by imposing a constraint on the MSE of the eavesdropper we also restrict the BER of the eavesdropper to be above a certain threshold. The resulting BER in the main channel, though, is also slightly degraded due to the secrecy restriction. We can also observe that the BERs that we can achieve when both receivers use ZF filters are lower than those when the legitimate receiver uses a ZF filter and the eavesdropper uses a Wiener filter (cf. Figures \ref{fig:BER1} and \ref{fig:BER2}). We argue that this seemingly counterintuitive behavior is due to the fact that in the scenario where the eavesdropper uses a Wiener filter instead of the ZF one, the transmitter cannot use all the available power.

\begin{figure}[!htbp]
    \begin{center}
        \mbox{\includegraphics[width=5.70in , height=3in]{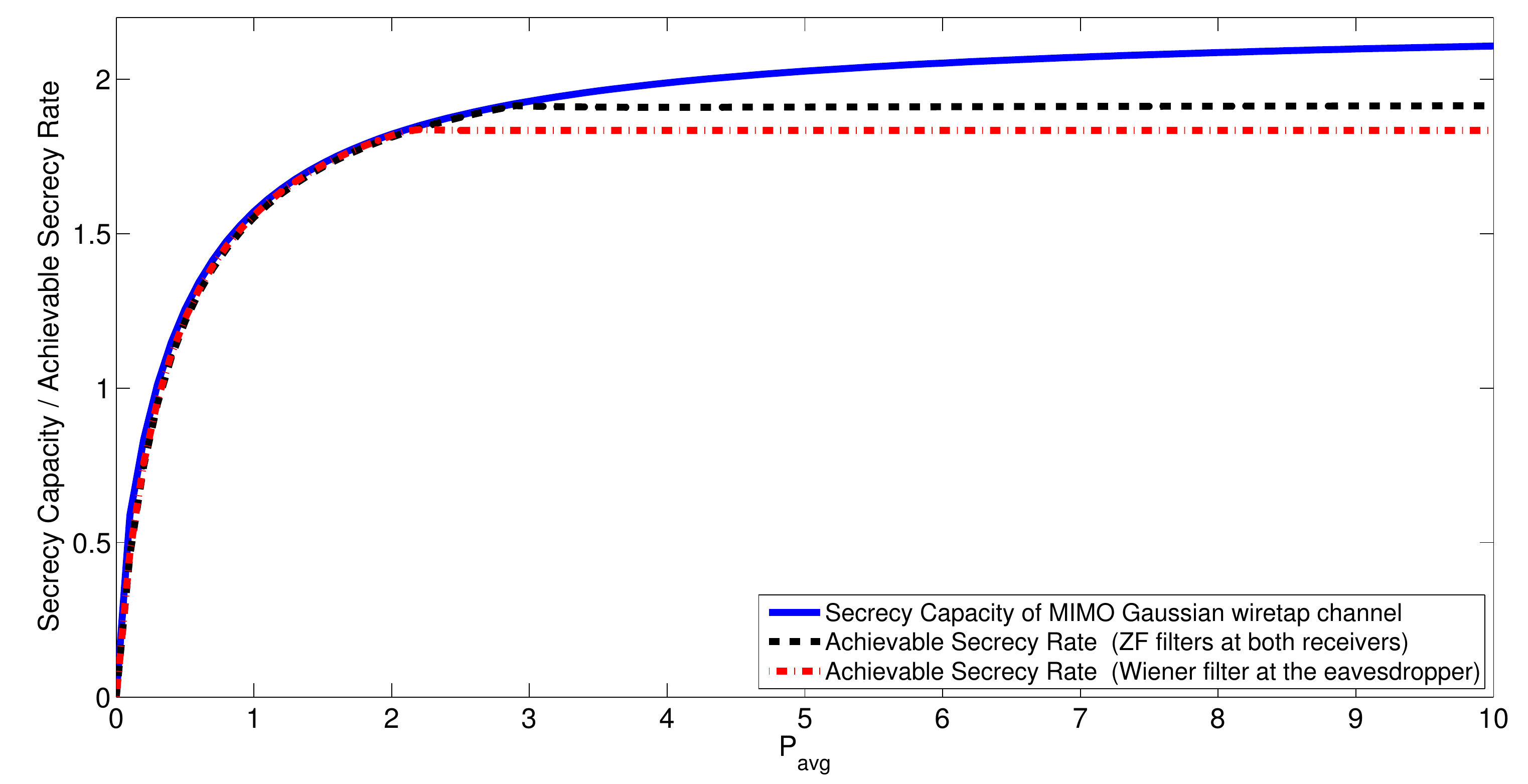}}
    \vspace{-0.5cm}
    \caption{Secrecy capacity of the MIMO Gaussian wiretap channel \emph{vs}. available power and achievable secrecy rate \emph{vs}. available power, for the optimal transmit filter design with ZF filters at both receivers and Wiener filters at the eavesdropper receiver ($\gamma = 0.5$).}    
    \label{fig:sec_rate}
    \end{center}
    \label{sec_rate}
\end{figure}

Finally, Figure~\ref{fig:sec_rate} compares the achievable secrecy rates yielded by our filter designs to the secrecy capacity of the MIMO Gaussian wiretap channel, which is given in~\cite{KhiWor10}. It is clear that the filter designs result in a loss of secrecy rate, which is more pronounced at high than at low available power levels, both for scenarios where the eavesdropper uses a ZF filter as well as scenarios where the eavesdropper uses a Wiener filter.

However, we note that our designs can be immediately realized in practice in order to impair the eavesdropper. In contrast, practical secrecy capacity achieving codes, which are known only for some special channels, have to be developed in order to achieve the secrecy capacity of the MIMO Gaussian wiretap channel.

\section{Conclusion}
\label{conclusions}
We have considered the problem of filter design with secrecy constraints in the classical wiretap scenario, where the objective is to conceive, subject to a power constraint, transmit and receive filters that
minimize the MSE between the legitimate parties whilst guaranteeing that the eavesdropper MSE remains above a certain threshold.

In particular, we have provided characterizations of the form of the receive and transmit filters for MIMO Gaussian channels, considering the situation where both receivers use Zero-Forcing filters or the eavesdropper uses a Wiener filter. We have also provided efficient computational procedures to design the optimal transmit and receive filters.

In particular, we have shown that the transmit filter designs are resilient to channel modeling errors as well as to the use of more powerful nonlinear receive filters, rather than the optimal linear Wiener filter, by the eavesdropper. We have also shown that the designs limit not only the eavesdropper MSE but also the error probability. 

We have also provided a framework to generalize this filter design problem from the scenario where all parties are assumed to know the exact state of the channel to scenarios where there is some channel uncertainty. This generalization is applicable not only to wireless systems subject to various channel state information regimes as well as to systems where there is uncertainty about the state of the eavesdropper channel. The generalization of the designs to cases where both receivers use optimal linear Wiener filters appear to be open in general.

\section*{Acknowledgment}
The authors thank the anonymous reviewers for their valuable comments and suggestions that significantly contributed to improving the quality of the paper. M. R. D. Rodrigues also thanks Kai-Kit Wong for very detailed comments about an earlier draft of the work.

\bibliographystyle{IEEEtran}
\bibliography{IEEEabrv,mybibfile}



\vspace{-0.5cm}

\vspace{-0.5cm}

\vspace{-0.5cm}

\vspace{-0.5cm}

\vspace{-0.5cm}

\vspace{-0.5cm}

\vspace{-0.5cm}

\vspace{-0.5cm}

\vspace{-0.5cm}

\vspace{-0.5cm}

\end{document}